\documentclass[useAMS,usenatbib]{mnras}
\usepackage{graphicx} 
\usepackage{subcaption}
\usepackage{amsmath} 
\usepackage{xcolor}
\usepackage{float}
\usepackage{caption}
\usepackage{placeins}
\usepackage{lipsum}
\usepackage{multicol}
\usepackage{soul}
\usepackage{amssymb}
\usepackage{ragged2e}
\usepackage{threeparttable}

\def\refjnl#1{{\rm#1}}
\def\aj{\refjnl{AJ}}                   % Astronomical Journal
\def\araa{\refjnl{ARA\&A}}             % Annual Review of Astron and Astrophys
\def\apj{\refjnl{ApJ}}                 % Astrophysical Journal
\def\apjl{\refjnl{ApJ}}                % Astrophysical Journal, Letters
\def\apjs{\refjnl{ApJS}}               % Astrophysical Journal, Supplement
\def\ao{\refjnl{Appl.~Opt.}}           % Applied Optics
\def\aap{\refjnl{A\&A}}                % Astronomy and Astrophysics
\def\mnras{\refjnl{MNRAS}}             % Monthly Notices of the RAS
\def\pasj{\refjnl{PASJ}}               % Publications of the ASJ
\def\nat{\refjnl{Nature}}              % Nature
\def\fcp{\refjnl{Fund.~Cosmic~Phys.}}  % Fundamental Cosmic Physics
\def\physrep{\refjnl{Phys.~Rep.}}   % Physics Reports
\def\be{\begin{equation}} 
\def\ee{\end{equation}}

\def\msun{{\Msun}}

\def\gsim{\lower.5ex\hbox{\gtsima}} 
\def\lsim{\lower.5ex\hbox{\ltsima}} \def\gtsima{$\; \buildrel > \over 
\sim \;$} \def\ltsima{$\; \buildrel < \over \sim \;$} \def\prosima{$\; 
\buildrel \propto \over \sim \;$} \def\gsim{\lower.5ex\hbox{\gtsima}} 
\def\lsim{\lower.5ex\hbox{\ltsima}} 
\def\simgt{\lower.5ex\hbox{\gtsima}} 
\def\simlt{\lower.5ex\hbox{\ltsima}} 
\def\simpr{\lower.5ex\hbox{\prosima}} \def\la{\lsim} \def\ga{\gsim} 
  
 \def\gtsima{$\; \buildrel > \over \sim \;$} 
\def\ltsima{$\; \buildrel < \over \sim \;$} 
\def\gsim{\lower.5ex\hbox{\gtsima}} 
\def\lsim{\lower.5ex\hbox{\ltsima}} 
\def\simgt{\lower.5ex\hbox{\gtsima}} 
\def\simlt{\lower.5ex\hbox{\ltsima}} 
\def\simpr{\lower.5ex\hbox{\prosima}} 
\def\la{\lsim} 
\def\ga{\gsim}

\def\msun{\,{\rm \Msun}}

\def\E3{{\cal E}_{\rm g}^{III}}

\def\msun{\rm M_\odot}

\def\Msun{\rm M_\odot}

\def\zsun{\rm Z_\odot}
\def\M*{M_*}

\def\Z*{Z_*}
\def\L*{L_*}
\def\muv{\rm M_{UV}}

\def\der{{\rm d}}

\newcommand\code[1]{\textsc{\MakeLowercase{#1}}}

\title[Dust in $z \sim 7$ REBELS LBGs]{The ALMA REBELS Survey: the dust content of $z \sim 7$ Lyman Break Galaxies} 
\author[Dayal et al.]{P. Dayal$^{1}$\thanks{p.dayal@rug.nl}, A. Ferrara$^2$, L. Sommovigo$^2$, R. Bouwens$^3$, P. A. Oesch$^{4,5}$, R. Smit$^6$, V. Gonzalez$^{7,8}$,
\newauthor S. Schouws$^3$, M. Stefanon$^3$, C. Kobayashi$^{9}$, J. Bremer$^{1}$, H. S. B. Algera$^{10}$, M. Aravena$^{11}$, 
\newauthor R. A. A. Bowler$^{12}$,  E. da Cunha$^{13,14}$, Y. Fudamoto$^{4, 15, 16}$, L. Graziani$^{17, 18}$, J. Hodge$^{19}$, H. Inami$^{10}$,  
\newauthor I. De Looze$^{20, 21}$, A. Pallottini$^{2}$, D. Riechers$^{22}$, R. Schneider$^{17, 23, 24, 25}$, D. Stark$^{26}$ and R. Endsley$^{26}$ \\
$^{1}$Kapteyn Astronomical Institute, University of Groningen, P.O. Box 800, 9700 AV Groningen, The Netherlands\\
$^{2}$Scuola Normale Superiore, Piazza dei Cavalieri 7, Pisa, 56126, Italy\\
$^3$Leiden Observatory, Leiden University, NL-2300 RA Leiden, Netherlands \\
$^4$Observatoire de Gen{\`e}ve, 1290 Versoix, Switzerland \\
$^5$ Cosmic Dawn Center (DAWN), Niels Bohr Institute, University of Copenhagen, Jagtvej 128, K\o benhavn N, DK-2200, Denmark \\
$^6$Astrophysics Research Institute, Liverpool John Moores University, 146 Brownlow Hill, Liverpool L3 5RF, United Kingdom \\
$^7$Departmento de Astronomia, Universidad de Chile, Casilla 36-D, Santiago 7591245, Chile \\
$^8$Centro de Astrofisica y Tecnologias Afines (CATA), Camino del Observatorio 1515, Las Condes, Santiago, 7591245, Chile \\
$^{9}$ Centre for Astrophysics Research, Department of Physics, Astronomy and Mathematics, University of Hertfordshire, College Lane \\
$^{10}$Hiroshima Astrophysical Science Center, Hiroshima University, 1-3-1 Kagamiyama, Higashi-Hiroshima, Hiroshima 739-8526, Japan \\
$^{11}$Nucleo de Astronomia, Facultad de Ingenieria y Ciencias, Universidad Diego Portales, Av. Ejercito 441, Santiago, Chile \\
$^{12}$Astrophysics, The Denys Wilkinson Building, University of Oxford, Keble Road, Oxford, OX1 3RH, United Kingdom \\
$^{13}$International Centre for Radio Astronomy Research, University of Western Australia, 35 Stirling Hwy, Crawley,26WA 6009, Australia \\
$^{14}$ARC Centre of Excellence for All Sky Astrophysics in 3 Dimensions (ASTRO 3D), Australia \\
$^{15}$Research Institute for Science and Engineering, Waseda University, 3-4-1 Okubo, Shinjuku, Tokyo 169-8555, Japan \\
$^{16}$National Astronomical Observatory of Japan, 2-21-1,  Osawa, Mitaka, Tokyo, Japan \\
$^{17}$Dipartimento di  Fisica, Sapienza, Universita di Roma, Piazzale Aldo Moro 5, I-00185 Roma, Italy \\
$^{18}$INAF/Osservatorio Astrofisico di Arcetri, Largo E. Femi 5, I-50125 Firenze, Italy \\
$^{19}$Leiden Observatory, Leiden University, NL-2300 RA Leiden, Netherlands \\
$^{20}$Sterrenkundig Observatorium, Ghent University, Krijgslaan 281 - S9, 9000 Gent, Belgium \\
$^{21}$Dept. of  Physics \& Astronomy, University College London, Gower Street, London WC1E 6BT, United Kingdom \\
$^{22}$I. Physikalisches Institut, Universit\"at zu K\"oln, Z\"ulpicher Strasse 77, D-50937 K\"oln, Germany \\
$^{23}$INAF/Osservatorio Astronomico di Roma, via Frascati 33, 00078 Monte Porzio Catone, Roma, Italy\\
$^{24}$Sapienza School for Advanced Studies, Viale Regina Elena 291, 00161 Roma Italy \\
$^{25}$Istituto Nazionale di Fisica Nucleare, Sezione di Roma1, Piazzale Aldo Moro 2, 00185 Roma Italy \\
$^{26}$Steward Observatory,  University of Arizona, 933 N Cherry Ave, Tucson, AZ 85721, United States \\
}
\begin{document} 
 
\date{} 

\maketitle

\begin{abstract}
{
We include a fully coupled treatment of metal and dust enrichment into the \code{Delphi} semi-analytic model of galaxy formation to explain the dust content of 13 Lyman Break Galaxies (LBGs) detected by the Atacama Large millimetre Array (ALMA) REBELS Large Program at $z\simeq 7$. We find that the galaxy dust mass, $M_d$, is regulated by the combination of SNII dust production, astration, shock destruction, and ejection in outflows; grain growth (with a standard timescale $\tau_0= 30$ Myr) plays a negligible role. The model predicts a dust-to-stellar mass ratio of $\sim 0.07-0.1\%$ and a UV-to-total star formation rate relation such that $log (\psi_{\rm UV})  = -0.05 ~[log (\psi)]^{2} + 0.86 ~log(\psi) -0.05$ (implying that 55-80\% of the star formation is obscured) for REBELS galaxies with stellar mass $M_* = 10^{9-10} \msun$. This relation reconciles the intrinsic UV luminosity of LBGs with their observed luminosity function at $z=7$. However, 2 out of the 13 systems show dust-to-stellar mass ratios ($\sim 0.94-1.1\%$) that are up to $18\times$ larger than expected from the {\it fiducial} relation. Due to the physical coupling between dust and metal enrichment, even decreasing $\tau_0$ to very low values (0.3 Myr) only increases the dust-to-stellar mass ratio by a factor $ \sim 2$. Given that grain growth is not a viable explanation for such high observed ratios of the dust-to-stellar mass, we propose alternative solutions.}
\end{abstract}


% *************************************************************
\section{Introduction}\label{intro}
% *************************************************************
Over the past decade instruments such as the Hubble Space Telescope (HST), Very large Telescope (VLT), Subaru and Keck have been used to assemble statistically large sample of high-redshift ($z \gsim 5$) Lyman break Galaxies (LBGs) at rest-frame ultra-violet (UV) wavelengths. These have provided excellent constraints on the key physical properties of early galaxies including the evolving UV luminosity function (UV LF), stellar mass function (SMF) and the redshift evolution of the star formation rate density and stellar mass density, to name a few \citep[see reviews by e.g.][]{dunlop2013, stark2016, dayal2018}. 

However, the impact of dust, which absorbs UV and optical photons that are re-emitted at Far Infra-red (FIR) wavelengths and can severely impact the UV-detectability of galaxies \citep{draine2003}, remained an open question especially at $z \gsim 5$. The advent of instruments such as the Plateau de Bure Interferometer (PdBI) and the Atacama Large millimetre Array (ALMA) have opened up a new window on the dust content of such early star forming galaxies \citep[for a review see][]{hodge2020}. Indeed, FIR observations have now been used to estimate dust-to-stellar mass ratios ranging between $0.012-3\%$ for ``normal" star forming galaxies, with stellar masses $M_* \sim 10^{8.3-10.5}\msun$, at $z \gsim 7$ \citep[e.g.][]{watson2015, laporte2017, marrone2018, hashimoto2019,bakx2020, reuter2020, fudamoto2021, schouws2021}.  

These have been complemented by a range of theoretical models aimed at understanding the key dust processes and their resulting impact on the visibility of galaxies. A number of zoom-in simulations have been crucial in understanding the role of a range of interstellar medium (ISM) processes - such as dust growth and dissociation, Supernova (SN) shock destruction, ISM dust grain growth and gas drag effects - on the dust distribution and its grain-size distribution in individual galaxies \citep{bekki2015, aoyama2017, mckinnon2018}. These have been supplemented by hydrodynamic simulations have both been post-processed citep{dayal2011, mancini2015, narayanan2018, wilkins2018, ma2019, vogelsberger2020, vijayan2021} and coupled with dust models \citep{li2019, graziani2020, kannan2021} to obtain the dust masses and resulting attenuation relations for high-redshift galaxies. Finally, a number of semi-analytic models have been constructed to study the impact of different processes on the dust content of high-$z$ galaxies \citep{popping2017, vijayan2019, triani2020}.

A key issue in determining the dust masses of early galaxies is that the observed FIR continuum emission is characterised by key two quantities - the dust temperature ($T_d$) and the dust mass ($M_d$). Unless multi-band dust measurements are available \citep[see e.g.][]{faisst2020, bakx2021}, these two quantities are degenerate, requiring an assumption on the dust temperature in order to infer the associated dust mass. This has led to two classes of explanations for the exceedingly high dust-to-stellar mass ratios seen for $z \gsim 7$ galaxies: the first focuses on invoking extremely fast grain growth in the ISM \citep[e.g.][]{mancini2015, michalowski2015}. The second is that the (luminosity-weighted) dust temperatures are, in fact, significantly higher -- up to $\approx 90$ K \citep[][]{behrens2018, sommovigo2020} -- than the usually assumed values of $\sim 35-40$ K \citep[see also][]{shen2021, vijayan2021}. Given that the FIR luminosity scales as $L_{\rm FIR}\propto M_d T_d^6$, this reduces the inferred dust mass by $2.1-2.5$ orders of magnitude, thus removing the need to invoke extreme grain growth rates that appear to be problematic at high-redshifts \citep{ferrara2016}. 

In order to build a larger sample of dusty star forming galaxies at high-$z$, an ALMA large program (the Reionization Era Bright Emission Line Survey; REBELS; PI: Bouwens) is underway. REBELS focuses on studying 40 of the brightest galaxies at $z \gsim 6.5$ over a 7 deg$^2$ area which are being scanned for both bright ISM cooling lines (such as those from [CII] 158 $\mu$m and [OIII] 88 $\mu$m) and dust continuum emission, as detailed in \citet{rebels_flagship}. This survey has already serendipitously revealed two dusty star forming galaxies at $z \sim 7$ \citep{fudamoto2021} in addition to yielding a sample of 13 (continuum and CII detected) $z \sim 7$ LBGs (Inami et al. 2022, in prep; Schouws et al. 2022, in prep. ) with $M_* \sim 10^{8.8-10.6}\msun$ (Stefanon et al. 2022, in prep; Topping et al. 2022, in prep.) and dust masses of $M_d \sim 10^{6.8-7.5} \msun$ (Sommovigo et al. 2022, in prep.); the associated dust-to-stellar mass ratios range between $\sim 0.2-1.1\%$. 

In this work, our goal is to: (i) study the key processes determining the dust content of  high-redshift ($z \gsim 7$) LBGs; and (ii) explore the impact of different ISM grain-growth timescales on the dust masses and UV-observability of such galaxies. To this end, we augment our \code{Delphi}\footnote{{\bf D}ark Matter and the {\bf e}mergence of ga{\bf l}axies in the e{\bf p}oc{\bf h} of re{\bf i}onization } semi-analytic model with a detailed treatment of chemical and dust enrichment in the ISM of LBGs. Although similar in spirit to the semi-analytic models noted above \citep[e.g.][]{popping2017, vijayan2019,triani2020}, the key strengths of this work lie in the fact that: (i) it uses a minimal number (two) of mass- and redshift-independent free parameters to model the key physics of early galaxies; (ii) contrary to the other models which have been calibrated against low-redshift ($z \sim 0-3$) data, our model has been calibrated against all available data sets for $z \gsim 5$ galaxies including the evolving UV LF and SMF. Once calibrated against these, our model also reproduces the observed redshift evolution of the star formation rate density and stellar mass density, to note a few; and (iii) this is the only semi-analytic model that includes the latest state-of-the-art yields from Type Ia SN (SNIa), Type II SN (SNII) and Asymptotic Giant branch (AGB) stars from \citet{kobayashi2020} so far. This yield set can reproduce the observations not only for oxygen but also for most of all stable elements (up to uranium) self-consistently.

We adopt a $\Lambda$CDM model with dark energy, dark matter and baryonic densities in units of the critical density as $\Omega_{\Lambda}= 0.691$, $\Omega_{m}= 0.308$ and $\Omega_{b}= 0.049$, respectively, a Hubble constant $H_0=100\, h\,{\rm km}\,{\rm s}^{-1}\,{\rm Mpc}^{-1}$ with $h=0.67$, spectral index $n=0.96$ and normalisation $\sigma_{8}=0.81$ \citep[][]{planck2016}. Throughout this work, we use a Salpeter initial mass function \citep[IMF;][]{salpeter1955} between $0.1-100\msun$ and a mass weighted solar metallicity value of $\zsun = 0.0122$ \citep{asplund2005}. Finally, we quote all quantities in comoving units, unless stated otherwise, and express all magnitudes in the standard AB system \citep{oke-gunn1983}.

We describe the \code{Delphi} model in Sec. \ref{model} before discussing the metal and dust models and their implementation in Sec. \ref{dust_model} and the rate evolution of the key dust processes in Sec. \ref{rate_ev}. We show the dust masses, dust-to-gas ratios and dust-to-metal mass ratios for $z\sim 7$ LBGs in Sec. \ref{dust_mass}. We then explore the effects of dust attenuation on the UV LF in Sec. \ref{dust_uvlf} and on the UV-to-total star formation rate (SFR) relation in Sec. \ref{dust_sfr} before concluding in Sec. \ref{conclusion}.

% *************************************************************
\section{Model}\label{model}
% *************************************************************
We start by briefly describing the \code{Delphi} model and interested readers are referred to \citet{dayal2014a} for complete details. \code{Delphi} uses a binary merger tree approach to jointly track the build-up of dark matter halos and their baryonic components (both gas and stellar mass). We start by building merger trees for 600 galaxies at $z=4.5$, uniformly distributed in the halo mass range of ${\rm log}(M_h/ \Msun)=8-14$, up to $z=40$ in equal time-steps of $\Delta t = 30$ Myr. This value is chosen so that all SNII from a given single stellar population explode within the same time-step \citep[SNII progenitors have lifetimes of $\lsim 28$ Myr;][]{padovani1993}. Each $z=4.5$ halo is assigned a co-moving number density by matching the $\der n / \der M_h$ value of the $z=4.5$ Sheth-Tormen halo mass function \citep[HMF;][]{sheth-tormen1999} and this number density is propagated throughout the merger tree of that halo. Thus, the resulting HMFs comply with the Sheth-Tormen one at all $z$. 

The very first progenitors of every $z=4.5$ halo, that mark the start of its assembly (``starting leaves"), are assigned an initial gas mass according to the cosmological baryon-to-dark matter ratio such that $M_g^{\rm i} = (\Omega_b/\Omega_m) M_h$. For halos that have progenitors, the halo mass is built via merging progenitors as well as smooth accretion of dark matter from the intergalactic medium (IGM). The initial gas mass in this case is the sum of the final gas mass inherited from its progenitors and that gained via smooth accretion making the \textit{Ansatz} that accretion of dark matter is accompanied by accretion of a cosmological fraction ($\Omega_b/\Omega_m$) of gas mass. A fraction of this initial gas mass is converted into stars with an effective star formation efficiency $f_*^{\rm eff} = \min[f_*^{\rm ej}, f_*]$, i.e. the minimum between (a) the efficiency that produces enough SNII energy to unbind the remainder of the gas ($f_*^{\rm ej}$), and (b) an upper maximum threshold, $f_*$ (see below).  

\begin{table*}
%\centering
\caption {A summary of the key physical parameters for two of the theoretical models explored in this work. For the model shown in column 1, we note the maximum instantaneous star formation efficiency (column 2), the fraction of SNII energy that couples to gas (column 3), the dust yield per SNII after SN shock processing (column 4), the fraction of warm ISM gas (column 5), whether the processes of dust destruction, astration and ejection have been included (columns 6-8), the product of the dust destruction efficiency and the fraction of SNII contributing to such shocks (column 9) and the dust grain growth timescale (column 10).    }
\centerlast
\begin{tabular}{|c|c|c|c|c|c|c|c|c|c|}
 \hline
  Model & $f_*$ & $f_w$ & $y_d [\msun]$ & $X_c$ & Dust destruction & Dust astration & Dust ejection & $f \epsilon$ & $\tau_0{\rm [Myr]}$ \\
   \hline
Fiducial & 8\% & 7.5\% & 0.5 & 0.5 & {\rm yes} & {\rm yes} & {\rm yes} & 0.03 &  30\\
Maximal & 8\% & 7.5\% & 1.0 & 0.5 & {\rm no} & {\rm yes} & {\rm no} & 0.03 &  0.3 \\
  \hline
  \end{tabular}
  \label{table1}
\end{table*} 

We compute the newly formed stellar mass at any $z$ as $M_*(z) = M_g^{\rm i} (z) f_*^{\rm eff}$; the corresponding SFR  is $\psi(z) = M_*(z)/\Delta t$. This star formation can impart the ISM with a total SNII energy given by $E_{SN} = f_w E_{51} \nu M_*(z)$, where $f_w$ is the fraction of SNII energy that couples to the gas, $E_{51} = 10^{51}{\rm erg}$ is the instantaneous energy produced per SNII and $\nu$ is the SNII rate per unit stellar mass formed; our chosen Salpeter IMF between $0.1-100\msun$ results in $\nu^{-1} = 134 \, {\rm \Msun}$. Further, at each step  we update the gas mass, including that lost in star formation and SN feedback. We also account for the gas return fraction ($R$) - this is the gas returned by exploding stars to the ISM for which we use the mass- and metallicity-dependent yields presented in \citet{kobayashi2020}. At the end of a redshift step $z$, the final gas mass is
\be
M_{g}^{\rm f}(z) = [M_g^{\rm i}(z)-M_*(z)] \bigg(1-\frac{f_*^{\rm eff}}{f_*^{\rm ej}}\bigg) + RM_*(z),
\ee 
and the ejected gas mass is
\be
M_{g}^{\rm eje}(z) = [M_g^{\rm i}(z)-M_*(z)] \bigg(\frac{f_*^{\rm eff}}{f_*^{\rm ej}}\bigg).
\ee

Each newly formed stellar population is assigned the metallicity of the ISM (computed as detailed in Sec. \ref{dust_model} that follows) at the previous time-step and assumed to have an age of $2$ Myr. These parameters are used in conjunction with the population synthesis code {\small STARBURST99} \citep{leitherer1999} to obtain the specific UV luminosity at $\lambda = 1500$\,\AA\, expressed as $L_{1500}\, [{\rm erg\, s^{-1} \, Hz^{-1}}]$\footnote{The specific UV luminosity includes both the stellar emission and nebular continuum with the latter contributing $\sim$15\% to the luminosity value.}. The total UV luminosity is the sum of this value plus the time-decayed UV luminosity from all older stellar populations in the galaxy.

Associated REBELS papers use a conversion between the UV luminosity and the SFR such that  
$ L_{\rm 1500} = \psi \kappa^{-1}$. While the REBELS collaboration uses a value of $\kappa = 7.1 \times 10^{-29}$ [${\rm \msun~ yr^{-1}~ erg^{-1}~ s ~Hz}$] based on a Chabrier IMF ($0.1-300\msun$), the paper inferring the UV SFR values (Ferrara et al. 2022, in prep.) uses a value of $\kappa = 4.45 \times 10^{-29}$ based on a Salpeter IMF ($1-100\msun$). Averaged over REBELS mass galaxies ($M_* \sim 10^{9-10}\msun$), we find a value of $\kappa = 8.9 \times 10^{-29}$ based on our Salpeter IMF ($0.1-100\msun$). In what follows, all inferred SFRs for REBELS sources are re-calibrated to our IMF and summarised in Table \ref{table2}. 

We note that our model contains only two mass- and redshift-independent free parameters to match to observations.  These are (a) the maximum (instantaneous) star formation efficiency of $f_* = 8\%$; and (b) the fraction $f_w (\approx 7.5$\%) of the SNII explosion energy  that is available to drive an outflow. These parameters have been tuned to simultaneously reproduce the observed SMF \citep[from][]{gonzalez2011, Duncan2014, Song2016}, and the UV LF at $z \sim 5-12$ \citep[from e.g.][]{castellano2010a, mclure2013, atek2015, finkelstein2015, bouwens2016, calvi2016, bowler2017,  livermore2017, ishigaki2018, oesch2018, harikane2021, bouwens2021}. While $f_*$ is crucial in determining the high-mass end of the SMF, $f_w$ determines the low-luminosity  and low-mass ends of the UV LF  and SMF, respectively \citep{dayal2014a}.  

%#################################################################
\subsection{Modelling the dust and metal contents of high-redshift galaxies}
\label{dust_model}
%#################################################################
The dust and metal contents of galaxies are inextricably interlinked. Dust and metals are produced by both supernovae and evolved AGB stars. However, AGB stars contribute only a few percent to the total dust mass for $z \gsim 5$ LBGs \citep[e.g.][]{dayal2010a, mancini2015, lesniewska2019} given their long evolutionary timescales. Whilst accounting for metals produced by both SNII and AGB, for the sake of simplicity, in this work, we assume that dust in $z \gsim 5$ LBGs is solely produced by SNII. We include the key processes of metals and dust production, astration, ejection and dust 
destruction and grain growth in the ISM. Given a lack of spatial information, we assume gas accreted from the IGM to be devoid of both metals and dust. However, as detailed in what follows, we do consider models without any ejection of metals and dust - in principle, their results would be comparable to a model wherein all ejected gas and metals stay in the circumgalactic medium and can be re-accreted at a later time-step. The evolution of the dust and the {\it gas-phase} metal masses as a function of time (or, equivalently, redshift) is described in detail in what follows\footnote{For the purposes of this work, we only track the total metal and dust masses, without tracking individual metal species.}. In our calculation, the {\it total} metal mass at any redshift is the sum of the gas-phase metal mass and the dust mass.

If a galaxy has no progenitors, we assume both the initial metal and dust masses to be zero i.e. $M_{\rm  Z}^i(z) =0$ and $M_d^i(z)=0$, respectively. On the other hand, for a galaxy with progenitors, the metal and dust masses at redshift $z$ are the sum of the metals and dust brought in by all the progenitors from previous time steps. At each time step, \code{Delphi} solves the following differential equation to calculate the change in the {\it gas-phase} metal mass using the instantaneous recycling approximation \citep[IRA;][]{tinsley1980} as: 
\begin{equation}
\frac{d M_{\rm Z}}{dt} = \dot M_{\rm Z}^{'\rm pro}  - \dot M_{\rm Z}^{\rm ast} + \dot M_d^{\rm des} -\dot M_{\rm Z}^{\rm eje} - \dot M_d^{\rm gro}.
\label{eq_met}
\end{equation}
The five terms on the RHS are the rates of: 
\begin{itemize}
\item {\it Metal production}: The first term, $\dot M_{\rm Z}^{'\rm pro}$ is the rate of gas-phase metal enrichment. It is calculated as $\dot M_{\rm Z}^{'\rm pro} = \dot M_{\rm Z}^{\rm pro} - \dot M_d^{\rm pro}$ where the first term on the right hand side shows the mass- and metallicity-dependent yields for stars between $0.1-50\msun$ using the results presented in \citet{kobayashi2020}; larger mass stars are assumed to collapse to black holes without producing any metals. The second term shows the rate at which  
dust condenses out of these metals, reducing the gas-phase metal content. This latter term is discussed in detail in what follows.

\item {\it Metal astration}: the assimilation of a homogeneous mixture of metals into stars which is expressed as $\dot M_{\rm Z}^{\rm ast} = [M_Z/M_g]\psi(t)$. 

\item {\it Dust destruction}: the dust mass that is destroyed in SNII shocks, $M_d^{\rm des}$, adds to the metal mass in the ISM. This term is detailed in what follows.

\item {\it Ejection}: of perfectly mixed metal-enriched gas in outflows such that $\dot M_{\rm Z}^{\rm eje} = [M_{\rm Z}/M_g] \dot M_g^{\rm eje}$. We assume the ejected gas is homogeneously ejected over the entire timestep so that $\dot M_g^{\rm eje} = M_g^{\rm eje}/\Delta t $. 

\item {\it Depletion}: this accounts for ISM metals lost to dust grain growth in the cold ISM. This term, $\dot M_d^{\rm gro}$, is also detailed in what follows.
\end{itemize}

Similarly, at each time-step, the dust mass is calculated solving the equation
\begin{equation}
\frac{d M_{d}}{dt} = \dot M_d^{\rm pro} - \dot M_d^{\rm ast} - \dot M_d^{\rm des} - \dot M_d^{\rm eje} + \dot M_d^{\rm gro}.
\label{eq_dust}
\end{equation}

The five terms on the RHS are the rates of the physical processes governing dust abundance: 
\begin{itemize}
\item[$\square$] \textit{Dust production}: by SNII is written as $\dot M_d^{\rm pro} = y_d \nu \psi(t)$ 
where $y_d = 0.5\, \msun$ is the adopted dust yield per SNII  \citep[e.g.][]{todini2001, bianchi2007}. This value is also in agreement with the dust yields ranging between $0.03-1.1\msun$ inferred from observations of SN in the local Universe \citep[e.g.][]{matsuura2015, temim2017, rho2018, priestley2020, niculescu2021}. However, a long standing question is how much of this dust can escape into the ISM, due to the destruction by the reverse shock \citep[see e.g.][]{bocchio2016, slavin2020}. 

\item[$\square$]\textit{Dust destruction}: by SNII shocks is expressed as 
\begin{equation} 
\dot M_d^{\rm des} = (1-X_c)\tau_{des}^{-1}\mathcal{D},
\end{equation}
where $\mathcal{D} = M_{d}/M_{g}$ is the dust-to-gas ratio and ($1-X_c$) is the mass fraction of warm ISM where dust can be destroyed. We use a fiducial value of $X_c=0.5$, based on recent high-$z$ galaxy simulation results \citep{Pallottini19}. The dust destruction timescale is 
\be
\tau_{des}^{-1} = f\,   \epsilon\,  \nu\,\psi M_s,
\ee
where $\epsilon$ is dust destruction efficiency by shocks for which we use a value of 0.2 \citep{mckee1989, seab1983}\footnote{The predicted range of $\epsilon$ ranges between 0.1-0.5, with the precise value depending on the ISM density and magnetic field strength.}, $f$ is the fraction of SNII that contribute to such shocks for which we use a value of 0.15 \citep{debennassuti2014} and $M_s {\rm(100\,  km\, s^{-1}) } = 6.8 \times 10^3 \, {\rm M_\odot}$ is the mass accelerated to $>100$ km ${\rm s^{-1}}$ by the SN  blast wave \citep{mckee1989, lisenfeld1998}. These values yield $f \epsilon = 0.03$ and $\dot M_d^{\rm des} = 0.76 \mathcal{D} \psi(t)$. 

\begin{figure*}
\begin{center}
\center{\includegraphics[scale=1.1]{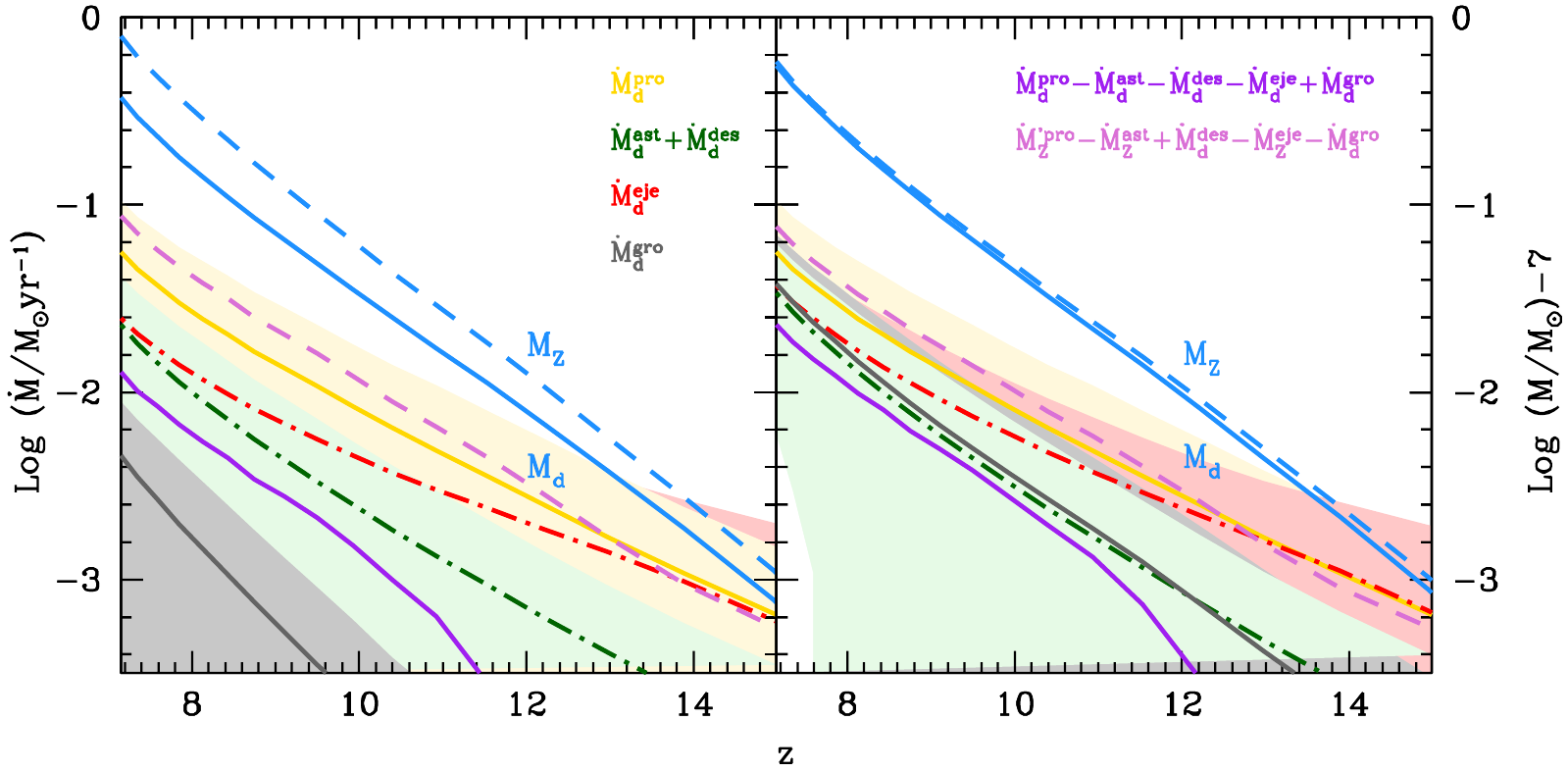}} 
\caption{Redshift evolution of the rates of the key processes determining the dust mass for $M_* = 10^{9-10}\msun$ galaxies at $z \sim 7$ for two different grain growth timescales (gray line): $\tau_0 =30$ Myr ({\it left panel}) and $\tau_0 =0.3$ Myr ({\it right panel}). The other lines show the (average values of the) different dust processes modelled: production (yellow), astration and destruction (dark green) and ejection (red); the shaded regions show the $1\sigma$ errors for each line. In each panel, the solid and dashed blue lines show the total dust mass and metal mass scaled down by 7 orders of magnitude, respectively. Finally, as marked in the right panel, the solid purple and dashed violet lines in both panels show the total rate of dust and metal mass assembly, respectively.  }
\label{fig_rates}
\end{center}
\end{figure*}

\item[$\square$]\textit{Dust astration}: is simply written as $ \dot M_d^{\rm ast} = \mathcal{D}\psi(t)$ where we assume astration of perfectly mixed gas and dust from the ISM.
\item[$\square$]\textit{Dust ejection}: with the same hypothesis of perfect mixture, is $\dot M_d^{\rm eje} = \mathcal{D}  \dot M_g^{\rm eje}$.

\item[$\square$]\textit{Grain growth}: occurs by accretion of heavy elements in the cold ISM component. An increase in the dust mass due to ISM growth must therefore be equalled by a decrease in the metal mass. This is modelled as \citep{dwek1998}
\be
\dot M_d^{\rm gro} = \bigg( 1 -\frac{M_d}{M_{\rm Z}+M_d}\bigg) \frac{X_c M_d}{\tau_{acc}},
\label{eqn_gr}
\ee 
where the first term ensures that the dust mass does not exceed the total metal mass and $\tau_{acc}$ is the grain growth timescale. A simple expression for the latter quantity is $\tau_{acc}= \tau_0 (Z/Z_\odot)^{-1}$ where $Z$ is the gas-phase metallicity ($=M_z/M_g$) and $\tau_0$ is a scaling timescale for which we explore values between $0.3-30$ Myr. Our $\tau_0$ values have been chosen in order to encompass the grain growth timescales explored in a number of other theoretical works \citep[e.g.][]{mancini2015, popping2017,vijayan2019, triani2020, graziani2020} up to the time difference between consecutive merger-tree steps.

\end{itemize}

We also calculate an (unphysical) upper limit to the dust mass ($M_d^{\rm max}$) considering the processes of dust production (assuming a yield of $y_d = 1\msun$), astration and growth assuming $\tau_0=0.3$Myr; dust is neither destroyed nor ejected in outflows in this case. This {\it gedankenexperiment} is termed the ``maximal dust mass model" and its implications are discussed in Secs. \ref{dust_mass} and \ref{dust_uv}.  

Using a single grain size and material density of $a= 0.05\, \mu m$ and $s= 2.25\, {\rm g\, cm^{-3}}$ appropriate for graphite/carbonaceous grains, respectively \citep{todini2001, nozawa2003}, this dust mass can be used to compute the ISM optical depth, $\tau_c$, to UV continuum photons as $\tau_c = 3 M_{d}[4 \pi r_g^2 a s]^{-1}$; we have assumed the extinction cross section of the grains to be $Q_{\rm ext}\approx 1$ at 1500\AA. Further, we have assumed dust and gas to be co-spatially distributed within the gas radius $r_g = 4.5 \lambda r_{\rm vir}$ \citep{ferrara2000}. We take the spin parameter $\lambda =0.04$ \citep{davis2009, dayal2018} where $r_{\rm vir}$ is the virial radius of the halo at the considered redshift. %For reference, $r_{\rm vir} \sim 15 ~ (60)$ physical Kpc for galaxies with $M_* \sim 10^9 ~ (10^{11})\msun$ at $z \sim 7$, resulting in $r_g \sim 2.7 ~ (10.8)$ physical Kpc\footnote{With its assumption of a homogeneous distribution of gas, stars and dust in the ISM, this represents an upper limit to the dust distribution radius; most of the dust would be expected to lie in the central regions (sub-kpc scales) with the density dropping off as one approaches $r_g$.    }.

This optical depth can be easily converted into a value for the escape fraction of continuum photons, $f_c$, by modelling the disk as a slab in which dust and stars are intermixed. This yields 
\be
f_c = \frac{1-e^{-\tau_c}}{\tau_c},
\ee
which for small optical depths is $\sim 1$. Note that $1-f_c$ can be also interpreted as the fraction of obscured SFR in the galaxy. 

The above equations have been implemented in \code{Delphi}, and solved self-consistently with  cosmological galaxy evolution. In the following, we will refer to the dust mass calculated with a grain growth timescale of $\tau_0 =30$ Myr as the {\it fiducial} model. The key model free parameters and their values are summarised in Table \ref{table1}.

% ******************************************
\subsection{Rate evolution of key dust processes}
\label{rate_ev}
% ******************************************
We start by clarifying the relative role of the different processes discussed in Eq. \ref{eq_dust} in shaping the dust content of early galaxies. To do this, we show the redshift evolution of the rate associated with each dust process, averaged over ``REBELS mass" galaxies with $M_* \sim 10^{9-10}\msun$ at $z\simeq 7$, in Fig. \ref{fig_rates}. In the {\it fiducial} scenario ($\tau_0 = 30$ Myr, left panel), such galaxies have a dust mass of $M_d \sim 10^{6.6}\msun$ by $z \sim 7$ with dust production dominating the dust mass assembly at all redshifts. Given their dependence on ${\cal D}$, the sum of the astration and destruction rates increase with $M_d$ (and hence, redshift) from being $\approx 26\%$ of the production rate at $z \sim 12$ to $\approx 40\%$ by $z \sim 7$. While the ejection rate also increases with time (given it is also $\propto \mathcal D$), its slope becomes shallower with decreasing redshift. This is because as halos grow in mass, the rate of gas and dust ejection decreases. Indeed, the ejection rate decreases from being about $\approx 67\%$ of the production rate at $z \sim 12$ to $\approx 44\%$ by $z \sim 7$. Finally, the ISM grain growth rate increases with metallicity, from about $ 1.4\%$ of the production rate at $z \sim 12$ to 8\% by $z\sim 7$. As seen, the total rate of dust growth shows a sharp decline at $z \gsim 12$ where the production rate is essentially balanced by the rates of dust astration, destruction and ejection. By $z \sim 7$, astration, destruction and ejection add up to about 84\% of the production rate, with grain growth playing a sub-dominant role. In this {\it fiducial} case, the redshift-evolution of the total dust mass for $M_* \sim 10^{9-10}\msun$ galaxies follows a roughly linear relation such that 
\be
{\rm log} M_d = -0.33z+8.9.
\ee 

\begin{figure*}
\center{\includegraphics[scale=1.01]{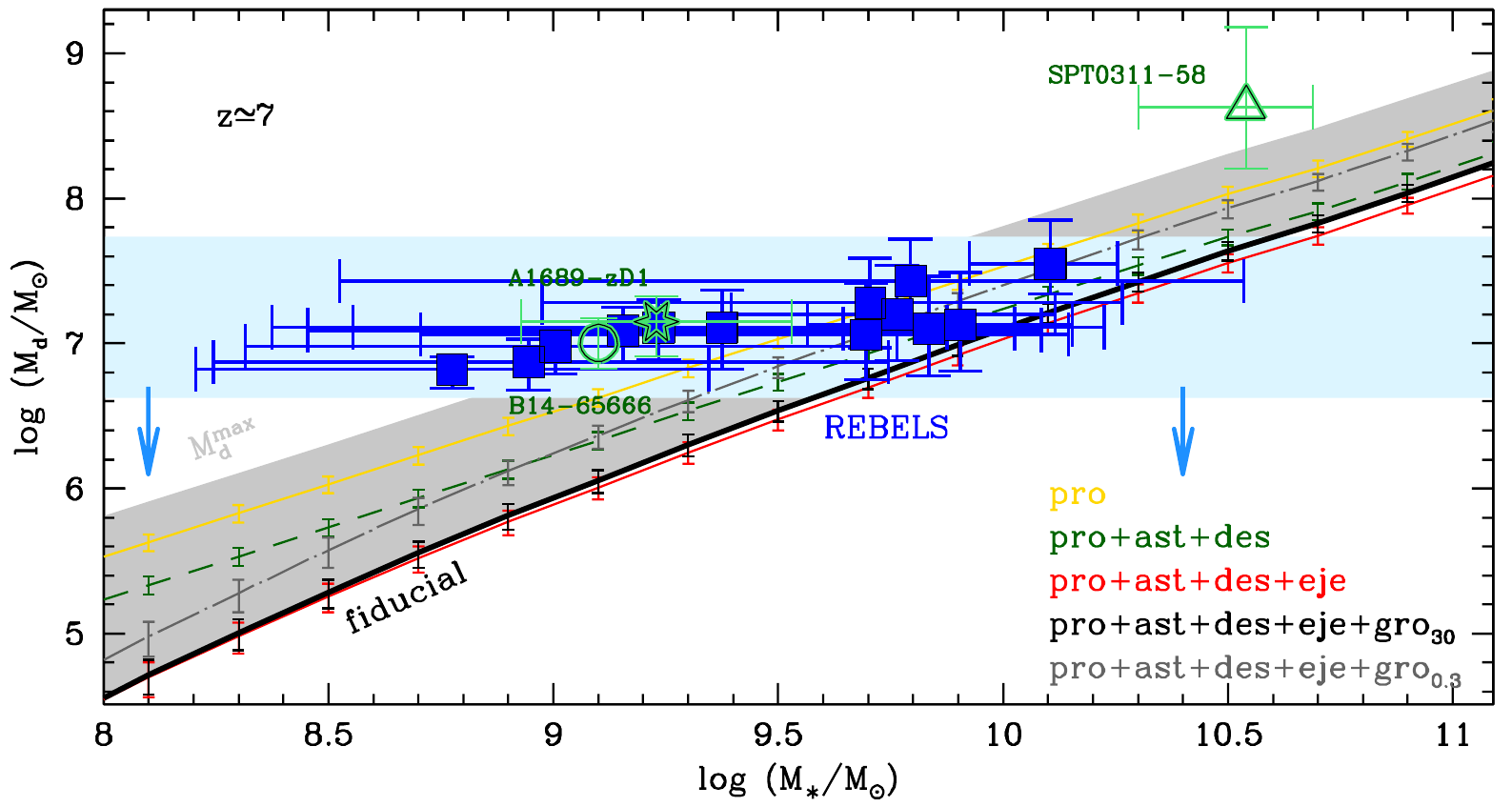}} 
\caption{The dust mass as a function of stellar mass at $z\sim 7$. Solid blue points show data from REBELS \citep[][Stefanon et al. 2022, in prep.]{rebels_flagship} where the stellar and dust masses have been re-scaled to a Salpeter IMF between $0.1-100 \msun$. The other points show observational data for A1689-zD1 (empty star) from \citet{watson2015} and \citet{bakx2021}, for B14-65666 (empty circle) from \citet{hashimoto2019} and for SPT0311-58 (empty triangle) from \citet{reuter2020}. The horizontal blue shaded strip shows the dust detection limits for the REBELS program using a non-detection flux limit of 66.49$\mu$Jy (Inami et al. 2022, in prep) assuming an average dust temperature that ranges between 30 and 80K (these show the upper and lower limits to the dust mass, respectively). Different lines (along with $1-\sigma$ error bars) show the progressive inclusion of the different dust-related physical processes discussed in Sec. \ref{dust_model} as in the label. The shaded gray area shows the range of dust masses demarcated by the {\it fiducial} model (lower limit) and the maximal dust mass model (upper limit; $M_d^{max}$).  }
\label{fig_mdms}
\end{figure*}

As seen from Fig. \ref{fig_rates}, ejection is the dominant mechanism responsible for decreasing the dust mass in the early assembly ($z \gsim 12$) of ``REBELS-mass" galaxies. Since our model assumes ejection of perfect mixed dust, gas-phase metals and gas, it results in the same fractional mass ejection of dust and metals. However, given the larger production rates (from both AGB and SNII) for metals and the fact that at any given time, dust destruction adds to the gas-phase metal mass, the rate of metal mass growth is less affected by ejection even at the highest redshifts, resulting in a high dust-to-metal mass. As these galaxies build-up their halo mass and the impact of ejection decreases at $z \lsim 10$, the rate of dust and metal growth start show similar slopes. This results in the dust mass-to-metal mass ratio decreasing with decreasing redshift. As shown in the same panel, $M_d/M_z$ decreases from $ \sim 60\%$ at $z \sim 12$ to $\sim 47\%$ by $z \sim 7$, with the metal mass being of the order $M_Z \sim 10^7 \msun$ by $z \sim 7$.

We then study the case where the grain growth timescale is a factor hundred shorter ($\tau_0 = 0.3$ Myr, right panel of the same figure); we note that in our formalism, the growth rate is regulated by the ratio of the dust and metal contents (Eqn. \ref{eqn_gr}). In this case too, the dust production term dominates at all $z$. Although the shorter grain growth timescales lead to a steeper increase in the total rate of dust growth and the associated dust mass with redshift as compared to the {\it fiducial} model, as a result of being modulated by the gas-phase metal mass, the final dust mass is only a factor 1.5 larger ($M_d \sim 10^{6.75}\msun$) by $z \sim 7$. As it might be expected, the steeper build-up of  dust mass also results in the astration+destruction and ejection rates showing a steeper slope (through their dependence on $\mathcal D$). Indeed, by $z \sim 7$, the astration+destruction and ejection rates are roughly $60\%$ of the production rate. Further, a larger fraction of metals condensing into dust in this model results in a dust-to-metal ratio that increases with decreasing redshift, from about $\approx 82\%$ at $z\sim 18$ to $\approx 95\%$ by $z \sim 7$. In this model, the total metal content is almost equally split between gas-phase ISM metals and those bound up into dust.  

To summarise, for $M_* \sim 10^{9-10}\msun$ galaxies at $z\sim 7$, the dust production rate dominates the dust build-up at all $z$ for both the grain-growth timescales considered here. In the {\it fiducial} model astration+destruction+ejection add up to $\sim$ 84\% of the production rate by $z \sim 7$, with grain growth playing a negligible role. However, invoking a model with $\tau_0= 0.3$ Myrs results in a steeper build-up of the dust mass resulting in a steeper rise in the astration+destruction and ejection rates; these processes each reach about 60\% of the production rate by $z \sim 7$.

% *************************************************************
\section{The dust content of high redshift galaxies}\label{dust_mass}
% *************************************************************
We now briefly describe how quantities such as the dust mass, stellar mass and star formation rates (UV and total) are derived for REBELS sources before these are compared to the model results. Interested readers are referred to Stefanon et al. 2022 (in prep.) and Topping et al. 2022 (in prep.) for complete details. The {\it fiducial} stellar masses for REBELS sources have been estimated from multi-wavelength photometry using a version of \textsc{Beagle} \citep{chevallard2016} that incorporates nebular emission \citep[both lines and continuum;][]{gutkin2016}. We have assumed a constant star formation history (SFH), a metallicity value of $0.2\zsun$, a Calzetti dust extinction law \citep{calzetti2000} and a Chabrier IMF between 0.1-300$\msun$ \citep{chabrier2003}. Additionally, Topping et al. 2022 (in prep.) have calculated the stellar masses adopting non-parametric SFHs\footnote{Using a Chabrier IMF, a metallicity of $0.2\zsun$ and a Small Magellanic Cloud (SMC) dust extinction law.. For the sources discussed in this work, the stellar masses derived by Topping et al. 2022 (in prep.) are, on average, 0.55 dex more massive than the values derived by Stefanon et al. (2022, in prep.). In what follows, we primarily compare to the {\it fiducial} stellar masses derived by Stefanon et al. (2022, in prep.) and comment on the differences with the Topping et al. 2022 (in prep.) results where appropriate.} Further, the dust masses for the REBELS sources are derived in Sommovigo et al. 2022 (in prep.) using the method presented in \citet{TdCIImetodo}. The central idea of this method is to use the CII luminosity ($L_{\rm CII}$) as a proxy for the total gas mass and therefore for the dust mass given a dust-to-gas ratio. Analytical formulas for both the dust-to-gas ratio and the CII-to-total gas conversion factors are provided in \citet{TdCIImetodo}. These authors also infer the associated dust temperature and the obscured star formation rates (i.e. in the FIR). The average dust temperatures for REBELS galaxies have a value of $T_d \sim 49$K, as independently inferred from the calculations presented in Sommovigo et al. 2022 (in prep.) and Ferrara et al. 2022 (in prep.) where the latter propose an analytic method solely based on the UV and FIR continuum information. Finally, the total SFR (i.e. the sum of the FIR and UV-deduced star formation rates) has been derived self-consistently with the dust temperature and mass for each galaxy in the sample using the model presented in Ferrara et al. 2022 (in prep.). All of these derived properties are reported in Table \ref{table2}. 

We now study the relation between the dust mass and stellar mass at $z \sim 7$  as shown in Fig. \ref{fig_mdms}. As expected, considering SNII dust production only results in a linear relation between the dust and stellar mass such that ${\rm log}\, M_d = {\rm log}\, M_*-2.42$ i.e. a dust-to-stellar mass ratio of about $0.38\%$. Including astration and dust destruction leaves the slope unchanged whilst leading to a decrease in the normalisation (by about 0.3 dex) resulting in the dust-to-stellar mass ratio decreasing to about $0.17\%$. Adding ejection leads to a steepening of this relation as low-mass halos ($M_* \lsim 10^9\msun$) lose a larger fraction of their gas (and dust) via outflows as compared to more massive systems. As seen from this plot, adding grain growth on timescales of $\tau_0=30$ Myr has only a slight effect on the relation such that for the {\it fiducial} model
\be
{\rm log}\, M_d = 1.15\,{\rm log}\, M_*-4.53,
\ee
for $M_* \sim 10^{8-11.5} \msun$ galaxies\footnote{Including lower-mass galaxies down to $M_* \sim 10^7 \msun$ introduces a steepening of this relation such that \be
{\rm log}\, M_d = -0.12\,{\rm log}\, M_*^2+3.63\,{\rm log}\, M_*-16.47.
\ee}. For REBELS mass galaxies, our {\it fiducial} model predicts a dust-to-stellar mass ratio that increases from $0.07$ to $0.1\%$ as the stellar mass increases from $10^9$ to $10^{10}\msun$. 

\begin{figure*}
\center{\includegraphics[scale=1.01]{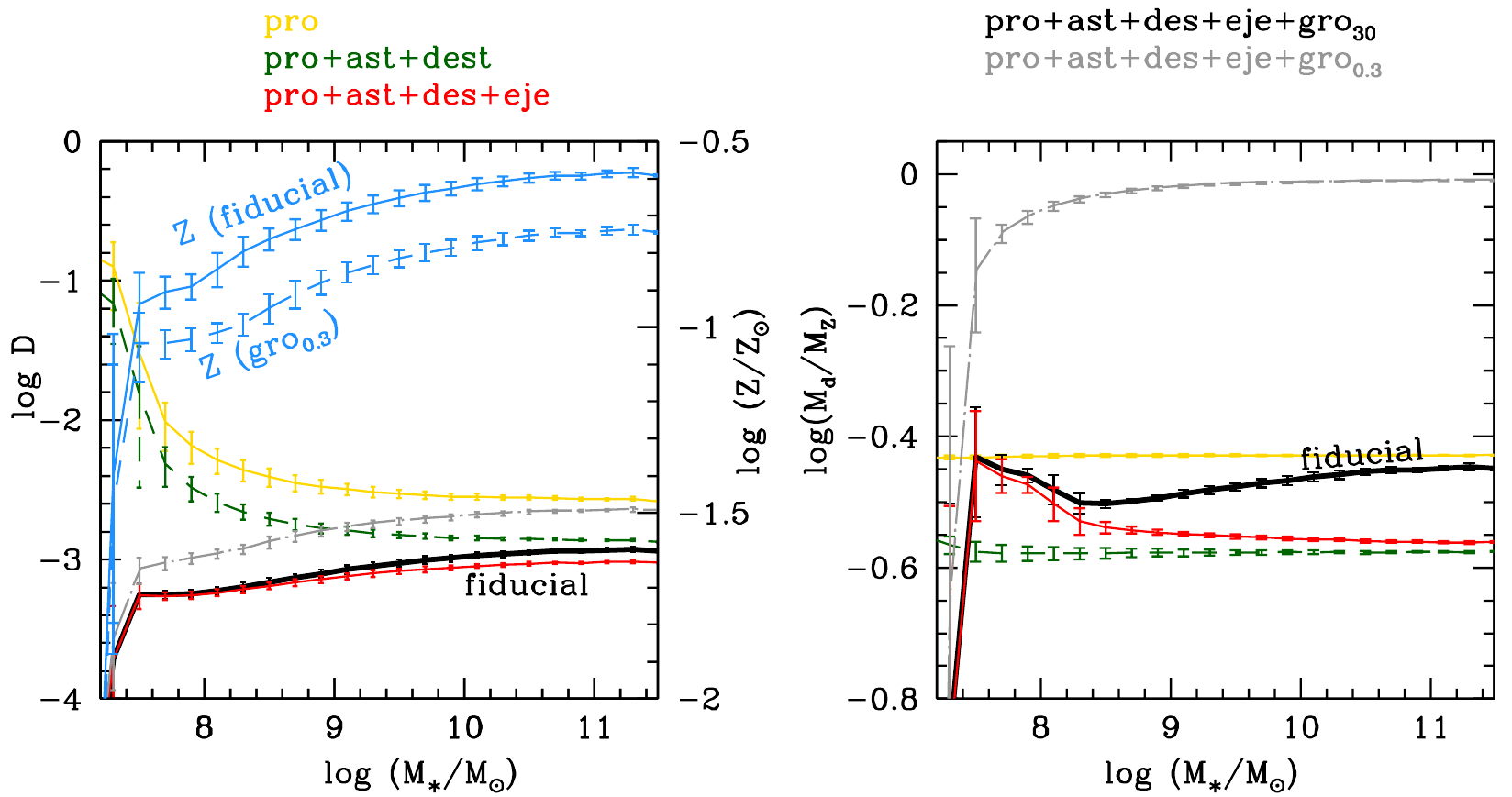}} 
\caption{As a function of the stellar mass at $z \sim 7$, for the different dust-related physical processes discussed in Sec. \ref{dust_model} and described in the label, we show: the dust-to-gas ratio ($\mathcal{D}$) and the metallicity normalised to solar units (blue lines) in the {\it left panel} and the dust-to-metal mass ratio ($M_d/M_Z$) in the {\it right panel}.}
\label{fig_dtg}
\end{figure*} 

Decreasing the grain growth timescale to $\tau_0 = 0.3$ Myr naturally results in the dust mass rising faster with stellar mass as compared to the {\it fiducial} model. This is because the increase in metallicity with stellar mass leads to shorter dust accretion timescales. However, even in this case, modulated by the gas-phase metal mass, the dust mass increases by, at most, a factor 2 ($\sim 0.3$ dex) for REBELS mass galaxies as compared to the {\it fiducial model} resulting in dust-to-stellar mass ratios of about $0.16-0.24\%$. It must be noted that even in this case, the dust masses still lie below the ``production" only model. Indeed, the upper limit is provided by the ``maximal dust mass" model wherein the dust masses are higher by about 0.8dex for $M_* \sim 10^{9-10}\msun$ galaxies as compared to the {\it fiducial} model - this model provides an upper limit to the dust-to-stellar mass ratio of about $0.63\%$. It must be cautioned that although they cannot be excluded\footnote{Physical arguments against rapid dust growth in early galaxies are reviewed in \citet{ferrara2016b}.}, growth timescales of the order of 0.3 Myr seem unlikely, if not unphysical. We end by noting that irrespective of the slope and normalisation, all the theoretical models studied here predict the dust mass to scale roughly linearly with the stellar mass.

We now discuss two interesting trends in terms of the data shown in Fig. \ref{fig_mdms}. Firstly, the observationally inferred dust masses for $z\sim 7$ galaxies seem to be essentially independent of the underlying stellar mass. This somewhat flat trend is consistent with the data points from REBELS \citep{rebels_flagship}, using either the fiducial stellar masses from Stefanon et al. (2022, in prep.) or those inferred by Topping et al. 2022 (in prep.), for B14-65666 \citep{hashimoto2019} and A1689-zD1 \citep{watson2015, bakx2021} as shown in this figure; the dust mass of SPT0311-58 \citep{reuter2020}, on the other hand, is consistent with the overall $M_d \propto M_*$ trend predicted by our estimates\footnote{However, we caution this source is composed of two individual galaxies at $z=6.9$, that are lensed by different amounts, complicating the picture.}. We then calculate the minimum dust mass that would be detectable by the REBELS program given its non-detection dust continuum flux limit of about 66.49$\mu$Jy. We find this minimum dust mass to have a value of about $10^{7.7} \, (10^{6.6}) \msun$ assuming a dust temperature of 30K (80K). This naturally biases our observations to only detecting galaxies with higher dust masses, especially for $M_* \lsim 10^{9.5} \, (10^{8.8})\msun$ systems as per the {\it fiducial} ({\it maximal} dust model, even assuming dust temperatures as high as 80K. 
The flatness of this relation can therefore be attributed to an observational bias introduced by the continuum detection limit of the REBELS program.

Secondly, and more crucially, within error bars the bulk of the REBELS observations are consistent with our {\it fiducial} model as seen from this figure. However, 2 of the lowest-mass galaxies (REBELS-19 and 39), with $8.7 < {\rm log} (M_*/\msun) < 9$, are clear outliers. These show dust masses and dust-to-stellar mass ratios ($\sim 0.94-1.1\%$) that are up to 18$\times$ higher for a given stellar mass as compared to the {\it fiducial} model (see also Table \ref{table2}); REBELS-14 is the third of the lowest-mass galaxies (reference stellar mass of $10^{8.73}\msun$) that is only consistent with the {\it fiducial} model if its stellar mass lies at the observationally-inferred upper limit of $M_* \sim 10^{9.7}\msun$. Matching REBELS-19 and 39 within error bars requires scenarios that lie between the ``no dust ejection" and ``maximal dust mass" models; the lowest-mass galaxy (REBELS-39) has a dust mass that lies between the ``production only" and ``maximal dust mass" models, even if its stellar mass is assumed to be at the upper limit predicted by observations; the above results remain qualitatively unchanged even if the higher stellar masses inferred by Topping et al. 2022 (in prep.) are used. Indeed, even ignoring any dust loss in astration, destruction and ejection, explaining the dust masses of these systems would require each SNII to produce $\sim 1.25-1.5\msun$ of dust (after SN shock processing; assuming a Salpeter $0.-100\msun$ IMF) which is higher than any observational estimate so far; using the stellar masses from Topping et al. 2022 (in prep.) would require each SNII to produce lower dust masses of $\sim 0.57-0.74\msun$, bringing them into agreement with observed SN dust yields (ranging between $0.03-1.1\msun$). Finally, we note that within error bars, the ``maximal dust mass" model also encompasses the dust masses inferred for A1689-zD1 \citep{watson2015, bakx2021}, B14-65666 \citep{hashimoto2019} and SPT0311-58 \citep{reuter2020}. 

We can also compute an upper limit to the dust mass from both production and grain growth for a galaxy of $M_* \sim 10^9 \msun$: assuming each SNII to produce $1 \msun$ of dust would yield $\sim 10^{6.9} \msun$ of dust using our chosen IMF. Further assuming all the gas-phase metals ($M_Z \sim 10^{6.4} \msun$) to have condensed into dust (i.e. resulting in a $M_Z=0$) would yield a total dust mass of $M_d \sim 10^7\msun$. This yields a dust-to-stellar mass ratio of $1\%$, assuming all the metals to have condensed into dust which is still below the inferred values for some of the observed sources (namely REBELS-19 and 39) and very close to the values inferred for A1689-zD1 and B14-65666. Alternatively, either (i) the dust mass deduced from the data could be overestimated, for example if the assumed temperature is too low as suggested by a number of theoretical works \citep[e.g.][]{behrens2018, liang2019, sommovigo2020}, or, more possibly, (ii) the stellar mass could be underestimated for these low-mass galaxies. Indeed, employing non-parametric SFHs, Topping et al., 2022 (in prep.) find the stellar masses of the lowest-mass galaxies ($M_* \lsim 10^{9.2}\msun$) to be higher by a factor 3-10. We compare our dust-to-stellar mass relation with those from a number of other semi-analytic models \citep[e.g.][]{popping2017, vijayan2019, triani2020} in Appendix \ref{comp_appendix}.

% This scenario is also rendered implausible by the fact that these sources are [CII] emitters (i.e. are not metal-free).

We then show the dust-to-gas-ratio ($\mathcal{D}$) as a function of the stellar mass in (the left panel of) Fig. \ref{fig_dtg}. As expected, this value is the highest for the production only model with $\mathcal{D} \sim 15\%$ for low-mass ($M_* \sim 10^{7.2} \msun$) galaxies given their low gas contents. For $M_* \gsim 10^9 \msun$ galaxies that are massive enough to retain most of their gas mass, the stellar (and dust) mass effectively scale with the gas mass resulting in $\mathcal{D}$ saturating at $\sim 0.27\%$. Including astration and destruction leads to a decrease in the dust-to-gas ratio by about 0.3 dex at $M_* \gsim 10^{8.5}\msun$ resulting in $\mathcal{D} \sim 0.14\%$. Preferentially decreasing the dust and gas mass contents of low-mass ($M_* \lsim 10^{7.5}\msun$) galaxies, including ejection completely changes the mass dependence of $\mathcal{D}$ by decreasing it by about 2.5 orders of magnitude at this low-mass end. At the high-mass ($M_* \gsim 10^9\msun$) end, ejection only decreases the amplitude of the dust-to-gas ratio by about 0.15 dex resulting in $\mathcal{D} \sim 0.1\%$. Including grain growth on a timescale of $\tau_0=30$Myr ({\it fiducial} model) has no appreciable impact on this relation. Similarly, including grain growth on a $\tau_0=0.3$ Myr timescale leads to only a small increase in $\mathcal {D}$ by about 0.3 dex compared to the fiducial model - in this case the value of $\mathcal{D}$ increases with stellar mass, reaching $\mathcal{D} \sim 0.2\%$ for $M_* \sim 10^{9-10}\msun$ galaxies. However, we reiterate that even in this case, the ``production only" model provides the upper limit to the dust-to-gas ratio. As shown in the same panel, in the {\it fiducial} model the metallicity for these galaxies increases from about 12\% to 26\% of the solar value as the stellar mass increases from $10^{7.5}$ to $10^{11.5}\msun$. As expected, a larger amount of metals saturate into dust assuming $\tau_0=0.3$Myr. This naturally results in a decrease in the metallicity values for all masses: in this case, the metallicity increases from 9-18\% for $M_* \sim 10^{7.5-11.5}\msun$ galaxies. Finally, for the {\it fiducial} model, we find a relation 
\begin{equation}
{\rm log}(\mathcal{D}) = 1.02 ~ {\rm log}(Z/\zsun) - 2.33. 
\end{equation}
This linear relation between the dust-to-gas ratio and the gas-phase metallicity is in accord with a number of previous observational and theoretical results \citep[e.g.][]{draine2007, leroy2011, remy-ruyer2013, li2019, hou2019}.

We then study the dust-to-metal ratio as a function of stellar mass (right panel of Fig. \ref{fig_dtg}). At $z \sim 7$, the dust-to-metal ratio has a constant value ($\sim 37\%$) in the production only scenario. Including astration and destruction naturally decreases this value, to $M_d/M_Z \sim 26\%$. Further including the process of ejection results an increase (of about 0.1 dex) in the dust-to-metal ratio for $M_* \lsim 10^{8}\msun$ halos whilst leaving the value largely unchanged at larger masses. This increase occurs because assuming perfect mixing (of metals and dust with gas) results in a larger amount of metals being ejected as compared to dust. Whilst including dust growth on a $\tau_0=30$Myr timescale hardly affects this ratio for low-mass galaxies ($M_* \lsim 10^{7.6}\msun$) given their low-metallicities, the dust-to-metal ratio increases with stellar mass for more massive halos; e.g. for $M_* \sim 10^{10}\msun$ galaxies, the value of $M_d/M_Z \sim 34\%$. Finally, including grain growth on a $\tau_0=0.3$Myr timescale leads to a steeper increase in the $M_d/M_Z$ value - for REBELS-type galaxies ($M_* \sim 10^{9-10}\msun$), $M_d/M_Z \sim 1$ is a factor 2.7 higher in this model compared to the {\it fiducial} model.

\begin{figure*}
\begin{center}
\center{\includegraphics[scale=1.01]{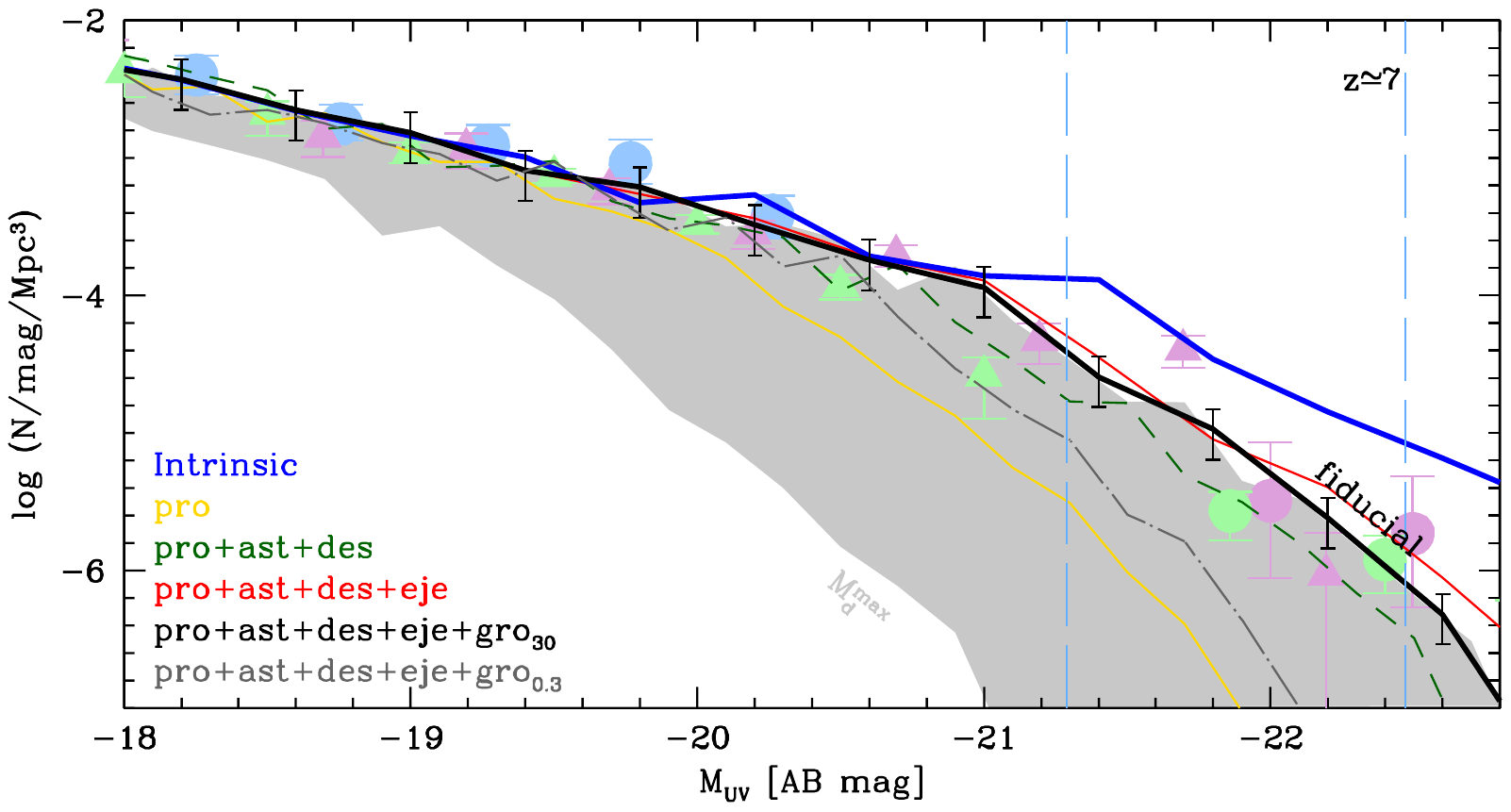}} 
\caption{The UV LF at $z \sim 7$. Points show observational data collected by different groups: \citet[][green triangles]{mclure2013}, \citet[][blue circles]{atek2015}, \citet[][green circles]{bowler2017}, \citet[][violet circles]{harikane2021} and \citet[][violet triangles]{bouwens2021}. The lines show the average UV LF for the dust processes noted, with the solid blue line showing the {\it intrinsic} UV LF. The shaded gray area shows the UV LF range demarcated by the {\it fiducial} model (upper limit) and the maximal dust mass model (lower limit; $M_d^{max}$). Finally, the dashed vertical lines show the range of UV magnitudes ($\muv \sim -22.47$ to $-21.29$) observed for REBELS sources.  }
\label{fig_uvlf}
\end{center}
\end{figure*}

%#################################################################
\section{Dust obscuration effects}
\label{dust_uv}
%#################################################################
We now discuss the observational implications of the presence of the predicted amount of dust in early galaxies, specially on their UV LF (Sec. \ref{dust_uvlf}) and the UV-to-total star formation rate relation (Sec. \ref{dust_sfr}).

%#################################################################
\subsection{The impact of dust obscuration on the UV LF}
\label{dust_uvlf}
%################################################################
As dust can heavily obscure UV radiation it is natural to base the analysis on the $z \sim 7$ UV LF, as shown in Fig. \ref{fig_uvlf}. As seen, the theoretical \textit{intrinsic} (i.e. unattenuated) UV LF starts over-predicting the number of bright galaxies with respect to observations at an absolute magnitude $\muv \lsim -21.2$. This implies that in these systems some dust attenuation is required, with the dust continuum FIR emission observed by REBELS confirming this hypothesis. 

However, only including SNII dust production results in a theoretical UV LF that is both lower in amplitude and steeper than the observed one in our model, cutting off at $\muv \sim -21.9$. The dust attenuation naturally decreases when the effect of astration and destruction are taken into account. As seen from the same figure, these two processes alone are almost sufficient to bring the UV LF into broad agreement with observations. Including ejection further increases the amplitude of the UV LF, that however, remains virtually unchanged once grain growth on a $\tau_0 = 30$ Myr is included. Thus, our {\it fiducial} model fits the observed UV LF at $z \sim 7$ for $\muv \sim -18$ to $-23$ extremely well. Assuming grain growth on a $\tau_0=0.3$ Myr timescale results in the model UV LF under-predicting the observations at $\muv \lsim -21.4$, showing a cut-off at $\muv \sim -22.1$. Finally, the ``maximal dust mass" model provides the lower limit to the UV LF, under-predicting the data for $\muv \lsim -19.2$ and showing a complete cut-off at $\muv \sim -21$. 

The large dust masses observationally inferred for $M_* \lsim 10^{9.5}\msun$ galaxies at $z \sim 7$ are therefore incompatible with the UV LF. This discrepancy could be reconciled if (i) such low-mass highly dusty galaxies are outliers that are not representative of the ``average" LBG population making up the UV LF; or (ii) not all of the dust mass contributes to UV attenuation. This second effect could be explained by a {\it spatial segregation} between dust and star forming regions (Inami et al. 2022, in prep. and Ferrara et al. 2022, in prep.) as has been found both in theoretical models \citep{mancini2016, behrens2018, liang2019, cochrane2019, sommovigo2020} and observations \citep{hodge2016,laporte2017, carniani2017, hashimoto2019} or most of the dust is diffused into the ISM with only a small fraction ($\sim 15\%$) attenuating the UV light from star forming regions; or (iii) the stellar masses have been under-estimated for these low-mass systems (see discussion in Topping et al. 2022, in prep.); or (iv) the dust temperatures have been under-estimated for low-mass systems, leading to an over-estimation of their dust content.

In any case, the UV LF provides valuable indication that the processes of production, astration, destruction and ejection are key in shaping the dust content, with grain growth only playing a minor role. 

%#################################################################
\subsection{The impact of dust obscuration on the UV-to-total star formation rate relation}
\label{dust_sfr}
\begin{figure*}
\begin{center}
\center{\includegraphics[scale=1.01]{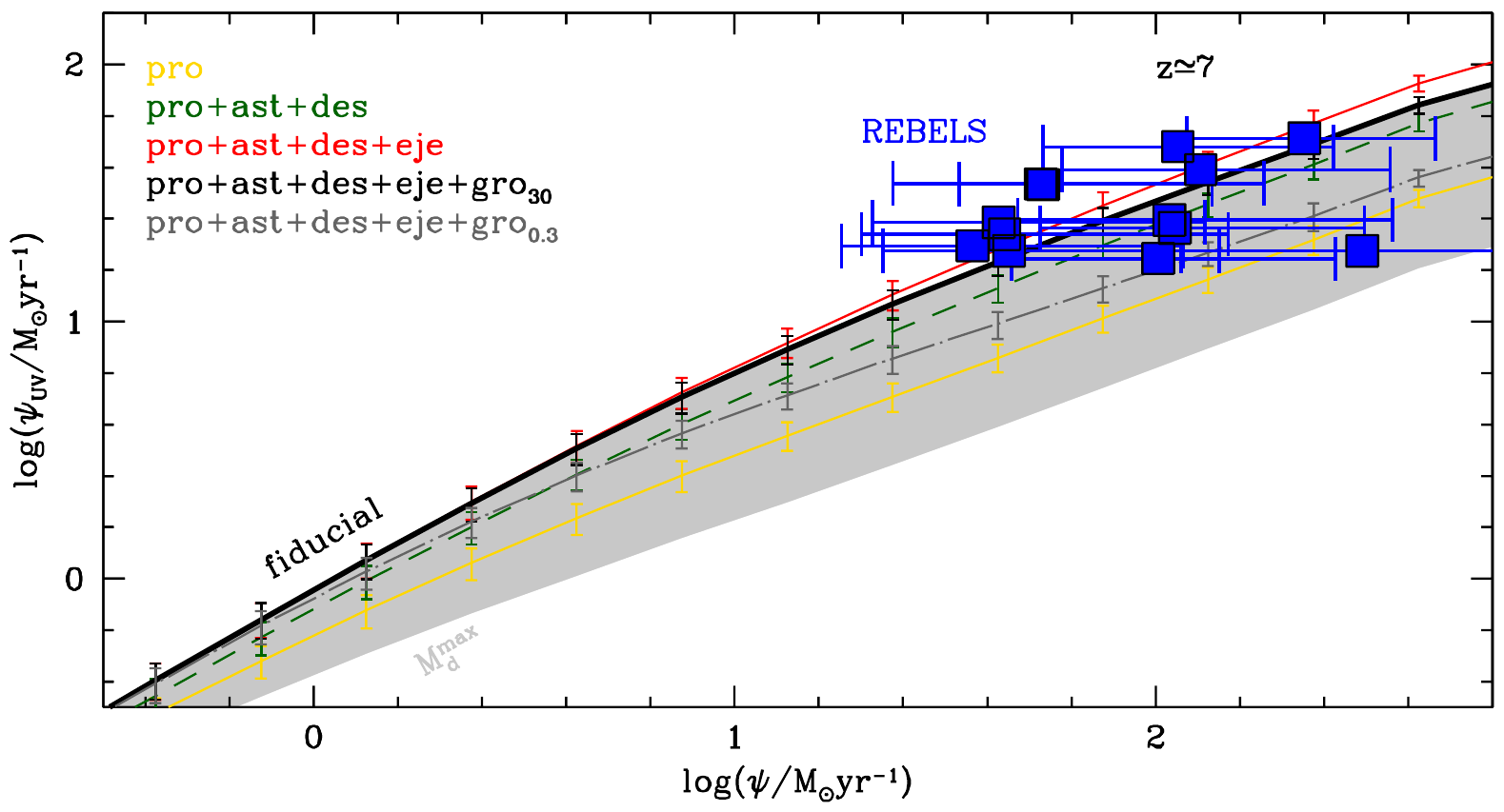}} 
\caption{The UV SFR ($\psi_{\rm UV}$) as a function of the total SFR ($\psi$) at $z \sim 7$. Filled squares show SFRs from the REBELS data (Ferrara et al. 2022, in prep.) re-scaled using a Salpeter IMF between $0.1-100 \msun$; lines are the predicted average values (along with $1-\sigma$ error bars) considering the different dust processes modelled, as noted. Finally, the shaded gray area shows the range of ${\rm SFR_{UV}}$ demarcated by the {\it fiducial} model (upper limit) and the maximal dust mass model (lower limit; $M_d^{max}$).
}
\label{fig_sfuvtot}
\end{center}
\end{figure*}

%################################################################
Finally, we discuss the relation between the SFR inferred from the UV luminosity of a galaxy ($\psi_{\rm UV}$), and the total intrinsic SFR ($=\psi$) in Fig. \ref{fig_sfuvtot}. As might be expected from the above discussions, for a given value of $\psi$, the value of $\psi_{\rm UV}$ progressively increases as the processes of dust astration, destruction and ejection are added to the ``dust production" only model. Including ISM grain growth on a 30Myr timescale naturally decreases $\psi_{\rm UV}$ slightly (by about 0.1 dex) with this {\it fiducial} model predicting a quadratic relation such that 
\begin{equation}
log (\psi_{\rm UV})  = -0.05 ~[log (\psi)]^{2} + 0.86 ~log(\psi) -0.05.
\end{equation}
This relation is valid for star formation rates ranging over 5 orders of magnitude, between $\sim 10^{-2.5} - 10^3 ~\msun {\rm yr^{-1}}$. For REBELS-type galaxies, with $\psi \sim 30-310 ~\msun {\rm yr^{-1}}$, our {\it fiducial} model predicts $\psi_{\rm UV} \sim 10-50 ~\msun {\rm yr^{-1}}$. While, within error bars, the bulk of the REBELS galaxies lie close to the fiducial model, there are a number of outliers (namely REBELS-19 and 25) that show $\psi_{\rm UV}$ values that a factor 2-3$\times$ smaller for a given $\psi$ value. These galaxies are more dust attenuated as compared to our {\it fiducial} model predictions which is in accord with the (up to 18$\times$) higher dust masses shown by some of the REBELS galaxies (specially REBELS-19) when compared to the {\it fiducial} model as detailed in Sec. \ref{dust_mass}. 

Compared to the {\it fiducial} model, the slightly higher dust masses (by a factor $\sim 2$) in the model with $\tau_0=0.3$ Myr result in a corresponding decrease in the UV SFR (that range between $8-30 ~\msun{\rm yr^{-1}}$). The results for this model therefore lie between the {\it fiducial}  and the ``production only" models. Finally, the ``maximal dust mass" model provides a lower limit to the $\psi_{\rm UV}$ for a given value of the total SFR. For REBELS galaxies, this model predicts $\psi_{\rm UV}$ values $\sim 5-13 ~\msun \, {\rm yr^{-1}} $ that are a factor 2-4 lower than those predicted by the {\it fiducial} model. We reiterate that with its extremely high extinction of UV light, the ``maximal dust mass" model is incompatible with the observed $z \sim 7$ UV LF.

\begin{figure*}
\begin{center}
\center{\includegraphics[scale=1.01]{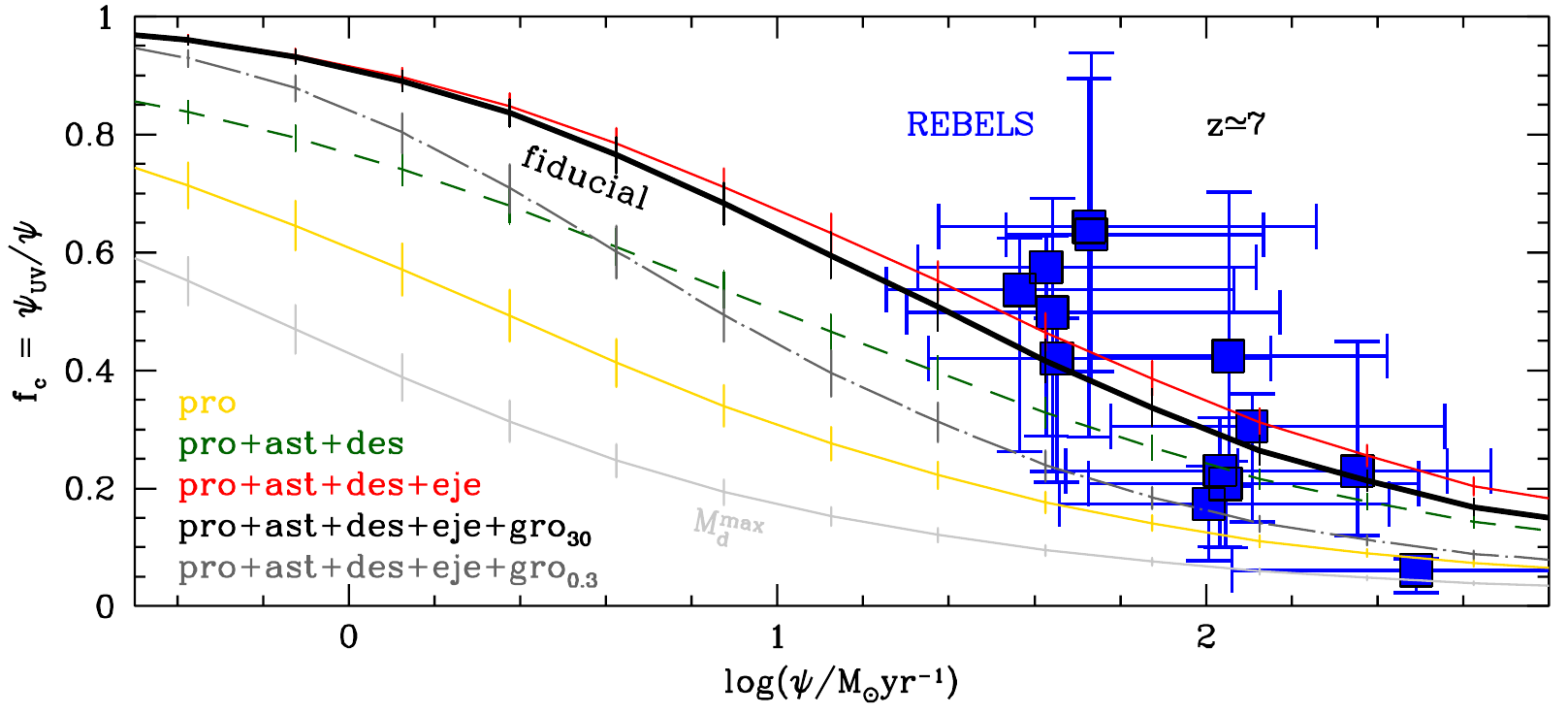}} 
\caption{The UV escape fraction ($f_c = \psi_{\rm UV}/\psi$) as a function of the total star formation rate for $z \sim 7$ galaxies. The different lines show the predicted average values (along with $1-\sigma$ error bars) considering the different dust processes modelled, as noted. The filled squares show the data from the REBELS program where the SFRs have been calculated as described in Ferrara et al. 2022 (in prep.) and re-scaled for a Salpeter IMF between $0.1-100 \msun$.
}
\label{fig_fc}
\end{center}
\end{figure*}

We now discuss the trend of $f_c$ as a function of the total star formation rate and the stellar mass, as shown in Fig. \ref{fig_fc}. We remind the reader that we have defined $f_c = \psi_{\rm UV}/\psi$. Starting with the ``production only" model, $f_c$ decreases by about an order of magnitude (from $\sim 0.75$ to $0.075$) as the total star formation rate increases from $\psi \sim 0.3$ to $630 ~\msun \, {\rm yr^{-1}}$ reflecting the increasing impact of dust attenuation in increasingly massive systems. Further, $\psi_{\rm UV} > 0.5$ only for galaxies with $\psi \lsim 2.4 ~\msun \, {\rm yr^{-1}}$ i.e. galaxies with higher total star formation rates will be increasingly suppressed in terms of their UV luminosity. The value of $f_c$ naturally increases when the processes of dust astration, destruction and ejection are added. Including these effects, the UV star formation rate dominates in galaxies with star formation rates as high as $\psi \sim 35 ~\msun \, {\rm yr^{-1}}$. These results change only slightly for the {\it fiducial} model where $f_c \sim 0.5$ for a slightly smaller value of $\psi \sim 25 ~\msun \, {\rm yr^{-1}}$. The model with grain growth on a 0.3 Myr timescale yields the steepest decrease of $f_c$ with $\psi$ as a result of the increase in dust mass in such massive systems. Finally, the ``maximal dust mass" model yields the lower limit to the unobscured star formation rate: in this model, $\psi_{\rm UV} \sim 0.5$, i.e. the UV SFR is obscured, for a total star formation rate value as low as $0.6 ~\msun \, {\rm yr^{-1}}$

We also see that the trend of $f_c$ shallows with increasing halo mass for all models - this is driven by the fact that the virial radius increases faster than the dust mass. For REBELS galaxies, $f_c$ only decreases by a factor of 2.25 from $\sim 45\%$ to $\sim 20\%$ as $\psi$ increases by about an order of magnitude from $\sim 40$ to $\sim 300 \msun ~ \rm{yr^{-1}}$, as seen in Fig. \ref{fig_fc}. The {\it fiducial} model therefore predicts that 55\% ($80\%$) of the SFR in the UV is obscured for galaxies with $\psi \sim 40 \, (300) \msun ~ \rm{yr^{-1}}$ which has enormous implications for the SFRD at such high-$z$. This is in accord with the results of Ferrara et al. 2022 (in prep.) who have independently derived attenuation values ranging between $28-91\%$ for the same galaxies based on their UV spectral slopes. 

With respect to comparison with observations, we limit these to results from the REBELS survey, whilst noting that similar results have been found at lower redshifts ($z \sim 4.4 - 5.8$) for about 118 galaxies in the rest-frame at 158$\mu$m as part of the Atacama Large Millimeter Array Large Program to INvestigate [CII] at Early times (ALPINE) survey \citep{fudamoto2020}. As seen, our results are in excellent agreement with the overall trend of $f_c$ decreasing with $\psi$ shown by the REBELS data. Furthermore, the bulk of these data points either lie on the {\it fiducial} model, or are consistent with it within error bars. The only two outliers in this case are also REBELS-19 and 25 as might be expected from the discussion above.
 
%#################################################################
\section{Conclusions and discussion}
\label{conclusion}
%#################################################################
In this work, we have included a fully coupled treatment of metal and dust enrichment into the \code{Delphi} semi-analytic model of galaxy formation. Our aim is to study the relative importance of the various processes (production, astration, destruction, ejection and ISM grain growth) regulating the dust content of high redshift LBGs. In addition to studying dust masses using two different ISM grain growth time-scales, $\tau_0=30$ Myr ({\it fiducial} model) and $\tau_0=0.3$ Myr, we explore a (clearly unphysical) ``maximal dust mass" model to provide an upper limit to the dust enrichment of such early galaxies. Our key findings at $z \sim 7$ are:
\begin{itemize}
\item In the {\it fiducial} model, dust production governs the dust content of $M_* \sim 10^{9-10} \msun$ galaxies over their entire lifetime; by $z \sim 7$. On average, for such galaxies, astration+destruction and ejection are each $\sim 40\%$ ($\sim 60\%$) as important as production for the {\it fiducial} ($\tau_0={\rm 0.3\, Myr}$) model. 

\item For the {\it fiducial} model, the dust and stellar mass are related as $M_d = 1.15\, {\rm log}\, M_* -4.53$. Predicting dust-to-stellar mass ratios $\sim 0.07-0.1\%$, this relation is in good agreement with the majority of the REBELS-detected points except for two sources (REBELS-19 and 39); these show dust-to-stellar mass ratios that are up to a factor 18 higher than the {\it fiducial} model - this result holds independently of the stellar masses inferred for REBELS galaxies (e.g. by Stefanon et al. 2022, in prep and Topping et al., 2022, in prep) that, on average, differ by 0.55 dex, depending on the SFH used. Crucially, due to the physical coupling between dust and metal enrichment, even decreasing $\tau_0$ to 0.3 Myr only increases the dust-to-stellar mass ratio by a factor $<2$. Hence, grain growth cannot be advocated to explain extremely high ratios.

\item In the {\it fiducial} model the dust-to-gas ratio is $\mathcal{D} \sim 0.1\%$ for $M_* \sim 10^{9-10}\msun$ REBELS-mass galaxies; for $\tau_0={\rm 0.3\, Myr}$ it only increases by a factor $\approx 2$, i.e. $\mathcal{D} \sim 0.2\%$.

\item The dust-to-metal mass ratio has a value $M_d/M_Z \sim 0.34$ in the {\it fiducial} model. This increases to $M_d/M_{\rm Z} \sim 1$ for the model with $\tau_0={\rm 0.3\,Myr}$ i.e. the total amount of metals produced by stars is equally split between the gas and the solid (grain) phase. 

\item Models not including dust attenuation largely over-predict the UV LF at $\muv \lsim -21.2$ \citep[see e.g.][]{bouwens2009, reddy2010}. The {\it fiducial} model yields instead results in excellent agreement with observations, and in close agreement with a model without any grain growth. However, grain growth with a 0.3 Myr timescale under-predicts the UV LF at all $\muv \lsim -21.4$.

\item The {\it fiducial} model predicts the UV SFR ($\psi_{\rm UV}$) and total SFR ($\psi$) to be related as $log (\psi_{\rm UV})  = -0.05 ~[log (\psi)]^{2} + 0.86 ~log(\psi) -0.05$ for $\psi \sim 10^{-2.6}-10^3~\msun {\rm yr^{-1}}$. 

\item The {\it fiducial} model predicts a UV escape fraction ranging between $20-45\%$ for REBELS galaxies i.e. 55\% (80\%) of the SFR in the UV is obscured for galaxies with $\psi \sim 40 \, (300) \msun ~ \rm{yr^{-1}}$. With its larger dust masses, the model with $\tau_0={\rm 0.3}$ Myr naturally predicts a larger attenuation fraction of $70\% \, (90\%)$ over the same SFR range.

\end{itemize}

\begin{table*}
%\centering
\caption {For the ID of the detected REBELS galaxies shown in column 1, we show the associated stellar mass (column 2), the luminosity weighted dust temperature (column 3), the dust mass (column 4), the UV SFR (column 5) and the total SFR (column 6).  }
\centerlast
\begin{tabular}{ |c|c|c|c|c|c|l|}
 \hline
  Object &  Redshift &  log($M_*/\msun)^a$ & $T_d [K]^b$ & log($M_d/\msun)^c$ & $\psi_{UV}$ [$\msun {\rm yr^{-1}}]^d$ & $\psi [\msun {\rm yr^{-1}}] ^e$  \\
   \hline
REBELS-05	& 6.496 &	$9.37^{+0.85}_{-1.0}$ & $44^{+15}_{-8}$ & $7.11^{+0.26}_{-0.29}$	& $18.8$ & $45.2^{+52.1}_{-23.5}$ \\
REBELS-08	& 6.749 &	$9.23^{+0.64}_{-0.68} $ & $54^{+16}_{-10}$ & $7.11^{+0.19}_{-0.22}$	& $23.1$	& $112.1^{+90.9}_{-53.1}$ \\
REBELS-12	& 7.349 &	$9.15^{+0.93}_{-0.70} $ & $54^{+16}_{-9}$ & $7.09^{+0.16}_{-0.22}$	& $39.0$	& $128.0^{+104.7}_{-59.8}$ \\
REBELS-14	& 7.084 &	$8.94^{+0.80}_{-0.70}$ & $56^{+13}_{-9}$ & $6.87^{+0.16}_{-0.19}$    & $47.9$    & $114.2^{+38.8}_{-54.0}$\\
REBELS-18	& 7.675 &	$9.70^{+0.56}_{-0.73} $ & $39^{+12}_{-7}$ & $7.28^{+0.31}_{-0.31}$	& $34.1$	& $46.5^{+27.7}_{-34.1}$ \\
REBELS-19	& 7.369 &	$9.00^{+0.69}_{-0.69} $ & $56^{+15}_{-10}$ & $6.98^{+0.13}_{-0.19}$	& $17.6$  	& $126.5^{+64.0}_{-45.3} $\\
REBELS-25	& 7.306 &	$10.1^{+0.15}_{-0.18} $ & $56^{+15}_{-14}$ & $7.55^{+0.30}_{-0.21}$	& $18.8$     & $310.8^{+214.3}_{-115.1}$ \\
REBELS-27	& 7.090 &	$9.90^{+0.25}_{-0.34} $ & $41^{+15}_{-9}$ & $7.13^{+0.36}_{-0.32}$	& $24.3$	& $40.7^{+46.2}_{-21.2}$ \\
REBELS-29	& 6.685 &	$9.83^{+0.19}_{-0.19} $ & $42^{+16}_{-10}$ & $7.10^{+0.36}_{-0.32}$	& $34.3$	& $53.3^{+74.0}_{-23.8}$ \\
REBELS-32	& 6.729 &	$9.76^{+0.35}_{-0.37} $ & $40^{+15}_{-9}$ & $7.20^{+0.34}_{-0.32}$	& $19.7$	& $31.2^{+42.6}_{-17.9}$ \\
REBELS-38	& 6.577 &	$9.79^{+0.74}_{-1.27} $ & $47^{+18}_{-10}$ & $7.43^{+0.29}_{-0.29}$	& $24.8$	& $105.8^{+149.7}_{-47.0}$ \\
REBELS-39	& 6.847 &	$8.77^{+0.57}_{-0.57} $ & $67^{+12}_{-9}$ & $6.82^{+0.09}_{-0.13}$	& $51.7$    & $210.5^{+9.4}_{-118.5}$ \\
REBELS-40	& 7.365 &	$9.69^{+0.45}_{-0.99} $ & $44^{+17}_{-10}$ & $7.06^{+0.35}_{-0.31}$	& $21.8$	& $44.8^{+60.9}_{-20.0}$ \\

  \hline
  \end{tabular}
  \begin{tablenotes}
 \raggedright{
{\tiny
\item $^a$ The stellar masses shown were derived in Stefanon et al. 2022 (in prep.) and have been re-scaled (up by 0.21 dex) so as to be consistent with a Salpeter IMF between $0.1-100\msun$.
\item $^{b, c}$ The dust masses and temperatures have been derived as detailed in Sommovigo et al. 2022 (in prep.) using a Salpeter $0.1-100\msun$ IMF.
\item $^d$ The UV SFR are the values derived in Ferrara et al. 2022 (in prep.) assuming a spherical dust distribution and a Milky Way extinction curve. These have been scaled up by 0.3 dex to be consistent with a Salpeter $0.1-100\msun$ IMF.
\item $^e$ The total SFR is the sum of the UV and FIR SFRs where the latter is calculated as $SFR_{IR}=L_{IR}/10^{10}$ as detailed in Sommovigo et al. 2022 (in prep.). 
}}
\end{tablenotes}
 \label{table2}
 \end{table*}

While the bulk of the REBELS observations are consistent with our {\it fiducial} model, there are 2 low-mass outliers (REBELS-19, 39) with $8.7 < {\rm log} (M_*/\msun) < 9.0$ that show dust-to-stellar mass values that are up to a factor 18$\times$ higher. This somewhat flat trend is also consistent with other data points for low-mass $z \sim 7$ galaxies \citep{watson2015, hashimoto2019}. Within error bars our unphysical ``maximal dust mass" model (that only includes production assuming a dust yield of $1 \msun$ per SNII, astration and grain growth on a $\tau_0=0.3$ Myr timescale) can reproduce the dust masses for the REBELS galaxies as well as for the other $z\sim 7$ dusty galaxies including A1689-zD1 \citep{watson2015, bakx2021}, B14-65666 \citep{hashimoto2019} and SPT0311-58 \citep{reuter2020}. However, such high dust masses result in an under-prediction of the UV LF for $\muv \gsim -19.5$. 

This tension can be resolved by four different possibilities: (i) such low-mass, highly dusty galaxies are outliers that are not representative of the ``average" LBG population making up the UV LF; (ii) not all of the dust mass observed contributes to UV attenuation either because dust and star forming regions are {\it spatially segregated} or a large fraction of dust is diffused into the ISM with only a small fraction contributing to attenuating UV light; (iii) the dust masses for these low-mass systems are over-estimated due to an under-estimation in the dust temperature; (iv) the stellar masses have been under-estimated, especially for the lowest mass systems observed by REBELS. Explanation (ii) is particularly relevant for REBELS-39 that shows a UV-to-total SFR relation in agreement with the {\it fiducial} model whilst having a dust mass that lies above this relation. REBELS-25 is an outlier that shows a larger attenuation of its UV SFR despite its dust mass being in perfect agreement with the {\it fiducial} relation. This might hint at dust being {\it preferentially clumped} around sites of young star formation. 

Finally, we end with a few caveats of the model. Firstly, we have assumed gas, metals and dust to be perfectly mixed in the ISM. Along the same lines, secondly, we have assumed a dust radius that is equal to the gas radius, and a homogeneous slab-like dust distribution within this. In principle, one might expect dust to be more concentrated in newly star forming regions and more dispersed into the ISM for older populations. Thirdly, we have ignored AGB contribution to the dust values inferred. However, as noted, this is expected to affect the inferred dust masses only slightly, i.e. at the $\sim 10\%$ level. Fourthly, we have neglected local over-densities of cold gas that might increase the ISM grain growth rate. Finally, we assume smoothly-accreted gas to be devoid of both metals and dust. This might be an under-estimation in the case of ``galactic fountains" i.e. when a part of the metal/dust-enriched gas ejected in at an earlier time is re-accreted onto the galaxy at a later stage. This is an effect that is already being included into our semi-numerical grid-based {\sc Astraeus} \citep{hutter2021} framework. Over the next years, a growing amount of ALMA data will be crucial in shedding light on a number of these outstanding issues. This includes properly sampled dust SEDs to address the dust temperature and the impact of the cosmic microwave background, matched-resolution ALMA/JWST (James Webb Space Telescope) imaging to directly address the patchy obscuration issue, adding more ``direct" ISM tracers than [CII] and higher spatial resolution observations to understand the UV and dust distributions (planned as part of the CRISTAL ALMA large program; PI: Herrera-Camus). 

%we could perhaps be more explicit in the last sentence, stating the "laundry list" of properly sampled dust SEDs to address Tdust (and the impact of the CMB), matched-resolution ALMA/JWST imaging to directly address the patchy obscuration issue, and adding more "direct" ISM tracers than [CII]

%
% ***************************************************************************
%\vspace{-0.8cm}
\section*{Acknowledgments} 
% ****************************************************************************
P. Dayal and J. Bremer acknowledge support from the European Research Council's starting grant ERC StG-717001 (``DELPHI"). P. Dayal also acknowledges support from the NWO grant 016.VIDI.189.162 (``ODIN") and the European Commission's and University of Groningen's CO-FUND Rosalind Franklin program and thanks Maxime Trebitsch for interesting discussions. A. Ferrara and A. Pallottini acknowledge support from the ERC Advanced Grant INTERSTELLAR H2020/740120. Any dissemination of results must indicate that it reflects only the author’s view and that the Commission is not responsible for any use that may be made of the information it contains; partial support from the Carl Friedrich von Siemens-Forschungspreis der Alexander von Humboldt-Stiftung Research Award is kindly acknowledged. R. J. Bouwens and M. Stefanon acknowledge support from TOP grant TOP1.16.057. R. Smit acknowledges support from a STFC Ernest Rutherford Fellowship (ST/S004831/1). S. Schouws acknowledges support from the Nederlandse Onderzoekschool voor Astronomie (NOVA). C. Kobayashi acknowledges funding from the UK STFC through grant ST/ R000905/1 \& ST/V000632/1. M. Aravena acknowledges support from FONDECYT grant 1211951, ``ANID+PCI+INSTITUTO MAX PLANCK DE ASTRONOMIA MPG 190030'', ``ANID+PCI+REDES 190194'' and ANID BASAL project FB210003. E.d.Cunha gratefully acknowledges support from the Australian Research Council Centre of Excellence for All Sky Astrophysics in 3 Dimensions (ASTRO 3D), through project number CE170100013. Y. Fudamoto acknowledges support from NAOJ ALMA Scientific Research Grant number 2020-16B. J. Hodge gratefully acknowledges support of the VIDI research program with project number 639.042.611, which is (partly) financed by the Netherlands Organisation for Scientific Research (NWO). H. Inami and H. S. B. Algera acknowledge support from the NAOJ ALMA Scientific Research Grant Code 2021-19A.
H. Inami acknowledges support from the JSPS KAKENHI Grant Number JP19K23462. I. De Looze acknowledges support from ERC starting grant DustOrigin 851622. R. Endsley acknowledges funding from JWST/NIRCam contract to the University of Arizona, NAS5-02015. This paper is based on data obtained with the ALMA Observatory, under the Large Program 2019.1.01634.L. ALMA is a partnership of ESO (representing its mem- ber states), NSF(USA) and NINS (Japan), together with NRC (Canada), MOST and ASIAA (Taiwan), and KASI (Republic of Korea), in cooperation with the Republic of Chile. The Joint ALMA Observatory is operated by ESO, AUI/NRAO and NAOJ.

\section*{Data Availability}
Data generated in this research will be shared on reasonable request to the corresponding author.
% **************************************************************************

%%%%%%%%%%%%%%%%%%%%%%%%%%%%%%
%\vspace{-0.8cm}
\bibliographystyle{mnras}
\bibliography{dust}

%\newpage
% **************************************************************************
\appendix
% **************************************************************************

% **************************************************************************
\section{Comparison of the dust-to-stellar mass relation with other semi-analytic models}
\label{comp_appendix}
% **************************************************************************
\begin{figure*}
\center{\includegraphics[scale=0.95]{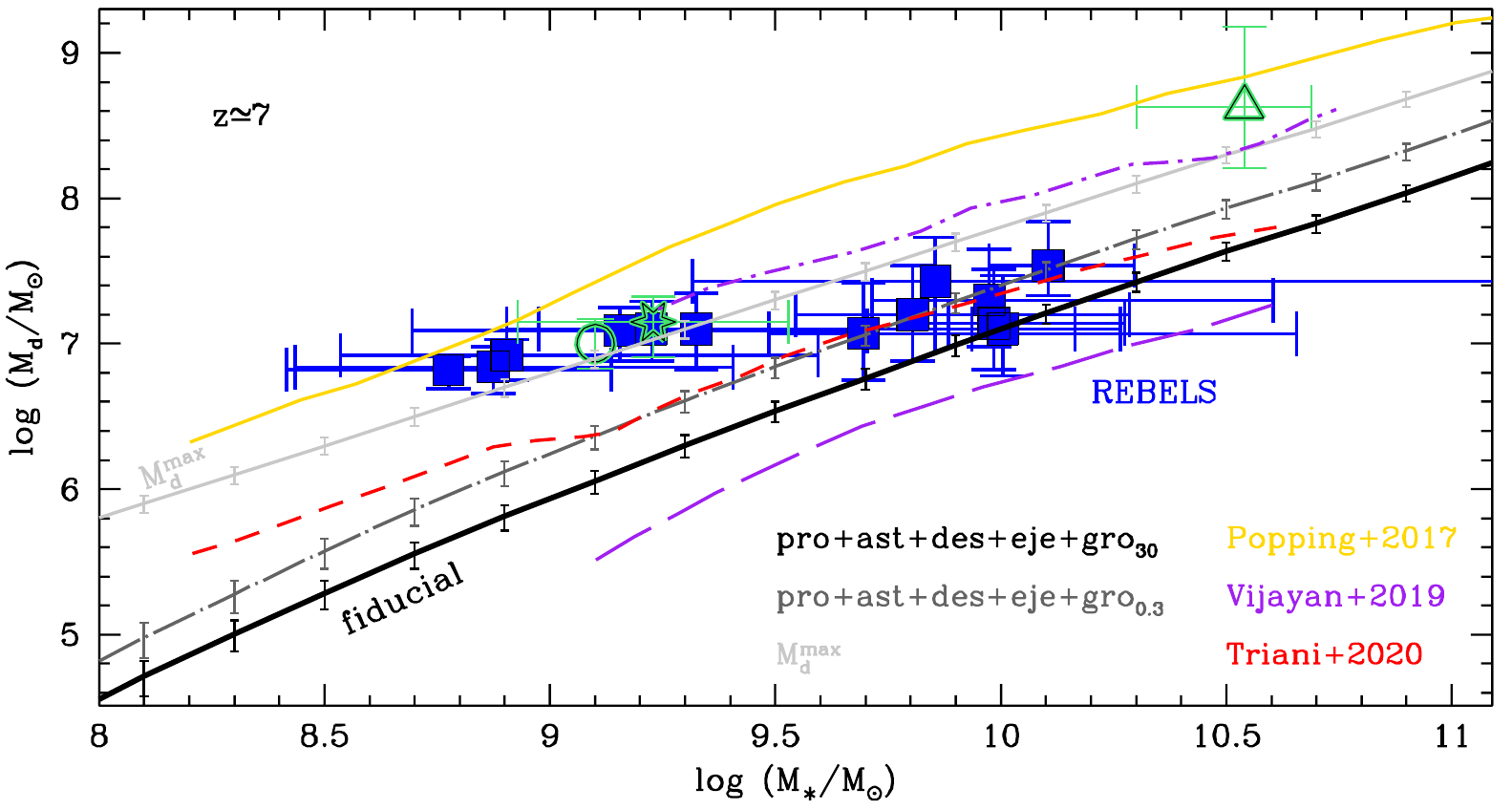}} 
\caption{The dust mass as a function of stellar mass at $z\sim 7$. Solid blue points show data from REBELS \citep{rebels_flagship} where the stellar and dust masses have been re-scaled to a Salpeter IMF between $0.1-100 \msun$. The other points show observational data for A1689-zD1 (empty star) from \citet{watson2015} and \citet{bakx2021}, for B14-65666 (empty circle) from \citet{hashimoto2019} and for SPT0311-58 (empty triangle) from \citet{reuter2020}. The solid black, dot-dashed gray and solid gray lines (along with $1-\sigma$ error bars) show results from this work for the {\it fiducial} model, a grain growth timescale of 0.3 Myr and the ``maximal dust mass" model, respectively. These are compared to the results from three other semi-analytic models: the solid yellow line shows the fiducial model from \citet{popping2017}, the violet dashed and dot-dashed lines show the fiducial model and upper limits from \citet{vijayan2019} and the red line shows the fiducial results from \citet{triani2020}.  }
\label{fig_mdms_comp}
\end{figure*}

In addition to being a key output of our model, the dust-to-stellar mass relation is one of the key observables for the REBELS program. We now compare our results to those from a number of semi-analytic models including the Santa Cruz model \citep{popping2017}, \textsc{L-Galaxies} \citep{vijayan2019} and \textsc{Dusty Sage} \citep{triani2020} that are able to simulate a statistically significant number of galaxies covering the REBELS stellar mass range ($M_* \sim 10^{9-10}\msun$). While all these models include (varying) prescriptions of the key physical process of gas cooling, star formation, SN feedback, chemical enrichment and dust (formation, astration, destruction, ejection, accretion), we caution they have been base-lined against low-redshift ($z \sim 0$) data as compared to our model that has been base-lined against all available observables at $z \gsim 5$. 

The Santa Cruz model \citep{popping2017} predicts the largest dust mass values for a given stellar mass. Finding a dust-to-stellar mass ratio that increases from $\sim 1\%$ to $4\%$ as $M_*$ increases from $10^9$ to $10^{10}\msun$, this model severely exceeds the average dust-to-stellar mass ratio of about $0.28\%$ found both by our model and by the REBELS program for $M_* \sim 10^{9.5-10}\msun$ galaxies. Indeed, this model sits at the upper limit for all the observed dust masses so far at $z \sim 7$. As discussed in Sec. \ref{dust_uvlf}, if such high dust masses were to be representative of the entire LBG population, the UV LF would be severely under-predicted due to dust attenuation unless this was compensated by intrinsically higher star formation rates. Physically, these high dust masses are possibly driven by the high production rate density in this model as well as the fact that they allow smooth accretion of metal- and dust-rich gas. Interestingly, although the amplitude is higher, the slope from this model is in good agreement with ours.

We then show both the fiducial as well as the maximal model results from the \textsc {L-Galaxies} model \citep{vijayan2019} that bracket the observed dust-to-stellar mass range. These authors assume all dust to be destroyed in major mergers in their fiducial model while we allow dusty mergers. Despite their different prescriptions for all key processes of galaxy formation, their fiducial model is only slightly lower than ours (by a factor 2.5) with a very similar slope. Their maximal model (that assumes saturated grain growth and no destruction) also lies very close to our ``maximal dust mass" model that also ignores dust destruction and assunes grain growth on a 0.3 Myr timescale compared to their $5 \times 10^4$ yr.

Finally, the results of the \textsc{Dusty sage} model (that also uses the Dwek grain growth model but an ISM grain growth timescale of 0.4 Myr) lie very close to our model that includes all the key dust process and where ISM grain growth takes place on a 0.3 Myr timescale. This is quite heartening, given their very different prescriptions for star formation, SN feedback and the fact that they too allow for accretion of metal and gas-rich IGM gas (the ``galactic fountain" model). While this model provides the best fit as compared to the other two for observed $M_* \gsim 10^{9.5}\msun$ galaxies, as in our model, it under-predicts the dust masses for lower-mass systems.

%Finally, while our results and those from REBELS have been shown for a $0.1-100\msun$ Salpeter IMF, these other works all use a Chabrier IMF. However, the uncertainties in the stellar masses are so large that any correction factor due to a different IMF would not change our results in the end. 

\label{lastpage} 
\end{document}